\documentclass[aps,prd,reprint,amsmath,citeautoscript,nofootinbib]{revtex4-1}
\usepackage[utf8]{inputenc}
\usepackage{bm}
\usepackage{amsmath,mathrsfs,amsfonts}
\usepackage{graphicx,afterpage}
\usepackage{subcaption}
\usepackage{amsmath,latexsym}
\usepackage{color,soul}
\usepackage{float}
\usepackage{silence}
\usepackage{physics}
\newcommand{\bea}{\begin{eqnarray}}
\newcommand{\eea}{\end{eqnarray}}
\newcommand{\be}{\begin{equation}}
\newcommand{\ee}{\end{equation}}
\def\slr{$SL(2,R)_L\times SL(2,R)_R$}
\def\sl2r{$SL(2,R)$}
\def\p{\partial}

\graphicspath{ {./images/} }
\begin{document}
\title{Hidden conformal symmetry of the rotating charged AdS black holes in quadratic $f$($T$) gravity}
\author{Canisius Bernard}
\email{canisius.bernard@usask.ca}
\author{Masoud Ghezelbash}
\email{masoud.ghezelbash@usask.ca}
\affiliation{Department of Physics and Engineering Physics, University of Saskatchewan, Saskatoon SK S7N 5E2, Canada}
\date{\today}
\begin{abstract}
The nonextremal Kerr black holes have been considered to be holographically dual to two-dimensional (2D) conformal field theories (CFTs). In this paper, we extend the holography to the case of an asymptotically anti--de Sitter (AdS) rotating charged black holes in $f$($T$) gravity, where $f(T) = T + \alpha T^2$, where $\alpha$ is a constant. We find that the scalar wave radial equation at the near-horizon region implies the existence of the 2D conformal symmetries. We note that the $2\pi$ identification of the azimuthal angle $\phi$ in the black hole line element, corresponds to a spontaneous breaking of the conformal symmetry by left and right temperatures $T_{L}$ and $T_{R}$, respectively.  We show that choosing proper central charges for the dual CFT, we produce exactly the macroscopic Bekenstein-Hawking entropy from the microscopic Cardy entropy for the dual CFT. These observations suggest that the rotating charged AdS black hole in $f$($T$) gravity is dual to a 2D CFT at finite temperatures $T_{L}$ and $T_{R}$ for a specific value of mass $M$, rotational, charge, and $f$($T$) parameters, $\Omega$, $Q$, and $\abs{\alpha}$, respectively.
\end{abstract}
\maketitle
\section{Introduction}
According to 't Hooft \cite{tHooft:1993dmi,Susskind:1994vu}, the combination of gravity and quantum mechanics requires the information in a three-dimensional (3D) world can be stored on a 2D projection, much like a hologram. This is well known as the holographic principle. This idea was inspired by the study of the entropy of the black holes. The essential outcome which is called the Bekenstein-Hawking formula \cite{Bekenstein:1973ur,Hawking:1974sw}, states that the entropy of a black hole is proportional to the area of the event horizon, $S=\text{Area}/4$. The actualization of this idea was discovered by Maldacena, in his well-known anti--de Sitter/conformal field theory (AdS/CFT) correspondence \cite{Maldacena:1997re}. The AdS/CFT duality relates a quantum field theory in $N-1$ dimensions and a quantum gravity theory in $N$ dimensions. The degrees of freedom of the CFT, live on the boundary of the AdS spacetime.\\
\indent The idea of AdS/CFT correspondence was extended to the case of extremal rotating black holes, namely, the Kerr/CFT correspondence which was proposed by Guica et al. \cite{Guica:2008mu}. The correspondence states that the physics of the extremal Kerr black holes, which are rotating with maximum angular velocity, can be described by a 2D CFT, living on the near-horizon region of the black holes. The correspondence was established by showing that one can microscopically reproduces the Bekenstein-Hawking entropy, using the CFT Cardy entropy formula. As one would expect, the Kerr/CFT correspondence is not only a peculiar property of extremal black holes but also nonextremal Kerr black holes. However, at the near-horizon region of the nonextremal Kerr black holes, one cannot indicate any conformal symmetries. In other words, the conformal symmetries are not the symmetries of the nonextremal Kerr black hole geometry (as they are for the case of the extremal Kerr black holes). However, it turns out that the ``hidden'' conformal symmetries can be revealed by looking at the solution space of the radial part of the Klein-Gordon equation, for a massless scalar probe in the near-horizon region of the  Kerr black holes \cite{Castro:2010fd}. In this case, the radial equation, can be written as the \slr~Casimir eigenequation. Subsequently, the Kerr/CFT correspondence can be established by matching the microscopic CFT Cardy entropy to the macroscopic Bekenstein-Hawking entropy of the Kerr black holes with general angular momentum and mass parameters. The correspondence has been studied for several black holes solutions, for instance, in Refs. \cite{Ghezelbash:2009gf,Ghezelbash:2009gy,Chen:2010bh,Chen:2010bd,Ghezelbash:2010vt,Ghezelbash:2012qn,Wang:2010qv,Chen:2010fr,Ghezelbash:2014aqa,Majhi:2015tpa,Astorino:2015lca,Astorino:2015naa,Astorino:2016xiy,Siahaan:2015xia,Siahaan:2018wvh}.\\
\indent The usual theory of gravity, based on the Riemannian geometry, has been extended through several gravity theories. One of them is the teleparallel gravity (TG) theory, where the Ricci scalar $R$, is replaced by the teleparallel torsion scalar $T$. Moreover, the TG has been generalized to $f$($T$) gravity by replacing the torsion scalar $T$, with an arbitrary function of $T$, such as $f$($T$). The charged black hole solutions in $f$($T$) gravity was found first in Refs. \cite{Gonzalez:2011dr,Capozziello:2012zj}. In \cite{Awad:2019jur}, Awad et al. find an asymptotically rotating charged AdS black hole solution, in quadratic $f$($T$) gravity, where $f(T) = T + \alpha T^2$. A very natural question to be asked, is ``Do we have any $f$($T$)/Hidden CFT correspondence?'', that we address in this article.\\
\indent The outline of this paper is organized as follows. In Sec. II, we review the $f$($T$)-Maxwell gravity theory as well as the rotating charged AdS black hole solutions, and its thermodynamics aspects. In Sec. III, we consider the massless Klein-Gordon wave equation, in the background of the rotating charged AdS black holes, in quadratic $f$($T$) gravity. In Sec. IV, we study the radial wave equation in the near-horizon region of the black holes, and rewrite it as the \slr~ Casimir equation. In Sec. V,  we find the central charges of the dual CFT by matching the Cardy entropy for the dual CFT to the macroscopic Bekenstein-Hawking entropy. Therefore, we present evidence that the rotating charged AdS black holes in quadratic $f(T)$ gravity, can be considered holographically dual to the CFT. In the final section, we summarize our results and address some future works. In this paper, we use the Planck units, in which $c=G=\hbar=k_{B}=1$.
\section{$f$($T$)-Maxwell gravity}
\subsection{Teleparallel gravity}
The basic variables in TG are tetrad fields ${e_a }^\mu$, where $a=\left(0,1,2,3\right)$ is the index of internal space and $\mu=\left(0,1,2,3\right)$ is the index of spacetime. The tetrad fields satisfy
\be
{e_a}^\mu {e^a}_\nu = \delta ^\mu _\nu,~{e_a}^\mu {e^b}_\mu = \delta ^b _a.
\ee
The tetrad fields are related to the spacetime metric and its inverse
\be
g_{\mu\nu}=\eta_{ab}{e^a}_\mu{e^b}_\nu,~g^{\mu\nu}=\eta^{ab}{e_a}^\mu{e_b}^\nu,
\ee
respectively, where $\eta_{a b}=\text{diag}\left(-,+,+,+\right)$ is the metric of 4D Minkowski spacetime. Also, it can be shown that $e=\det(e_{a\mu})=\sqrt{-g}$, where $g$ is the determinant of the metric. In TG, we use the Weitzenbock connection
\be
{W^\alpha}_{\mu\nu}={e_a}^\alpha \partial_{\nu} {e^a}_\mu=-{e^a}_\mu \partial_\nu {e_a}^\alpha,
\ee
to define the covariant derivative, by
\be
\nabla_\nu {e_a}^\mu =\partial_\nu {e_a}^\mu + {W^\mu}_{\rho \nu} {e_a}^\rho = 0.
\ee
The Weitzenbock connection is curvaturefree, but it has a non vanishing torsion
\be
{T^\alpha}_{\mu\nu}={W^\alpha}_{\nu\mu}-{W^\alpha}_{\mu\nu}={e_i}^\alpha\left(\partial_\mu {e^i}_\nu - \partial_\nu {e^i}_\mu\right).
\ee
We define the torsion scalar by
\be
T={T^\alpha}_{\mu\nu}{S_\alpha}^{\mu\nu},
\ee
where the superpotential tensor is
\be
{S_\alpha}^{\mu\nu}=\frac{1}{2}\left({K^{\mu\nu}}_\alpha+\delta^\mu_\alpha{T^{\beta\nu}}_\beta-\delta^\nu_\alpha{T^{\beta\mu}}_\beta\right).
\ee
We note that the contortion tensor $K_{\alpha\mu\nu}$ is given by
\be
K_{\alpha\mu\nu}=\frac{1}{2}\left(T_{\nu\alpha\mu}+T_{\alpha\mu\nu}-T_{\mu\alpha\nu}\right).
\ee
\subsection{Rotating charged AdS black holes}
In this paper, we consider a four-dimensional rotating charged AdS black hole solution in $f$($T$)-Maxwell theory with a negative cosmological constant where
\be \label{ft}
f\left(T\right)=T+\alpha T^2.
\ee
The dimensional negative parameter $\alpha$ is the coefficient of the quadratic term of the scalar torsion. The action of the $f$($T$)-Maxwell theory in 4D, for an asymptotically AdS spacetimes, is given by
\be \label{action}
\mathcal{S} = \frac{1}{{2\mathfrak{K}}}\int {{d^4}x\left| e \right|\left( {f\left( T \right) - 2\Lambda  - F{ \wedge ^ * }F} \right)}, 
\ee
where $\Lambda  =  - 3/{l^2}$ is the 4D cosmological constant, $l$ is the length scale of AdS spacetime. The constant $\mathfrak{K}$ in (\ref{action}) is related to the 4D Newton's gravitational constant $G_4$, by $\mathfrak{K}  = 2{\Omega _2}{G_4}$, where ${\Omega _2} = 2{\pi ^{3/2}}/\Gamma \left( {3/2} \right)$, is the volume of 2D unit sphere, and $\Gamma(3/2)=\frac{1}{2}\sqrt{\pi}$. In action (\ref{action}), $F = d\tilde \Phi$, where $\tilde \Phi  = {\tilde \Phi _\mu }d{x^\mu }$ is the gauge potential one-form.\\
\indent Varying action (\ref{action}) with respect to the tetrad fields and the Maxwell potential $\Phi_{\mu}$, one finds the field equations for gravity
\bea \nonumber
&&{S_\mu }^{\rho \nu }{\partial _\rho }Tf''\left( T \right)\\\nonumber 
&\quad& + \left[ {{e^{ - 1}}{e^a}_\mu {\partial _\rho }\left( {e{e_\alpha }^\alpha {S_\alpha }^{\rho \nu }} \right) - {T^\alpha }_{\lambda \mu }{S_\alpha }^{\nu \lambda }} \right]f'\left( T \right)\\\label{einstein}
&\quad& - \frac{{\delta _\mu ^\nu }}{4}\left( {f\left( T \right) + \frac{6}{{{l^2}}}} \right) =  - \frac{\mathfrak{K} }{2}{\mathcal{T_{\text{em}}}^\nu }_\mu, 
\eea
and the Maxwell's equations
\be\label{maxwell}
{\partial _\nu }\left( {\sqrt { - g} {F^{\mu \nu }}} \right) = 0.
\ee
respectively. In Eq. (\ref{einstein}), ${\mathcal{T_{\text{em}}}^\nu }_\mu={F_{\mu \alpha }}{F^{\nu \alpha }} - 1/4{\delta _\mu }^\nu {F_{\alpha \beta }}{F^{\alpha \beta }},$ is the energy-momentum tensor of the electromagnetic field. The rotating charged AdS black hole solution, is given by \cite{Awad:2019jur}
\bea\nonumber
d{s^2} &=& - A(r){\left( {\Xi dt - \Omega d\phi } \right)^2} + \frac{{d{r^2}}}{{B(r)}} \\\label{metric}
&\quad& + \frac{{{r^2}}}{{{l^4}}}{\left( {\Omega dt - \Xi {l^2}d\phi } \right)^2} + \frac{{{r^2}}}{{{l^2}}}d{z^2},
\eea
where the range of coordinates are given by $-\infty < t,z < \infty $, $0 \le r < \infty $ and $0 \le \phi < 2\pi $. In metric (\ref{metric}),  we have 
\be \label{Am}
A(r) = {r^2}{\Lambda _{eff}} - \frac{M}{r} + \frac{{3{Q^2}}}{{2{r^2}}} + \frac{{2{Q^3}\sqrt {6\left| \alpha  \right|} }}{{6{r^4}}},
\ee
\be\label{Bm}
B(r) = A(r)\beta(r), 
\ee
\be\label{para}
\beta(r) = {\left( {1 + Q\sqrt {6\left| \alpha  \right|} /{r^2}} \right)^{ - 2}},\\
\ee
\be
\Xi = \sqrt {1 + \frac{\Omega^2 }{{{l^2}}}},
\ee 
where ${\Lambda _{eff}} = \frac{1}{36\left| \alpha  \right|}$, and $M$, $Q$, and $\Omega$ are the mass parameter, the charge parameter, and the rotation parameter, respectively.  The parameter $\alpha$ cannot be zero, since the effective cosmological constant $\Lambda _{eff}$, and the metric functions $A(r)$ and $B(r)$ become singular. The gauge potential one-form {$\tilde \Phi$} is given by
\be
\tilde \Phi (r) =  - \Phi (r)\left( {\Omega d\phi  - \Xi dt} \right).
\ee
where $\Phi(r)= \frac{Q}{r} + \frac{{{Q^2}\sqrt {6\left| \alpha  \right|} }}{{3{r^3}}}$. We note that the torsion scalar $T$, for the black hole solution (\ref{metric}), is given by
\be\label{torsionscalar}
T(r)=\frac{{4A'(r)B(r)}}{{rA(r)}} + \frac{{2B(r)}}{{{r^2}}},
\ee
where $A(r)$ and $B(r)$, in Eqs. (\ref{Am}) and (\ref{Bm}).\\
\indent We notice that setting the rotational parameter $\Omega=0$, we find the static charged black hole configuration, as in Ref. \cite{Awad:2017tyz}. Moreover, turning off the mass parameter $M$ and $Q$, the metric (\ref{metric}) reduces to, the 4D AdS metric in an uncommon coordinate system.\\
\indent The horizons of the black holes are the positive roots of $A(r)=0$, among which, the outer one is denoted by $r_+$. The nonvanishing components of the contravariant metric tensor are given by
\bea\nonumber
{g^{tt}} &=& \frac{{{l^4}\left( {A(r){\Omega ^2} - {r^2}{\Xi ^2}} \right)}}{{A(r){r^2}{\left( {{\Xi ^2}{l^2} - {\Omega ^2}} \right)}^2}},~{g^{rr}} = B(r),~{g^{zz}} = \frac{{{l^2}}}{{{r^2}}},\\\nonumber
{g^{\phi \phi }} &=& \frac{{A(r){\Xi ^2}{l^4} - {r^2}{\Omega ^2}}}{{A(r){r^2}{\left( {{\Xi ^2}{l^2} - {\Omega ^2}} \right)}^2}},\\\label{contra}
{g^{t\phi }} &=& {g^{\phi t}} = \frac{{\Xi \Omega {l^2}\left( {A(r){l^2} - {r^2}} \right)}}{{A(r){r^2}{\left( {{\Xi ^2}{l^2} - {\Omega ^2}} \right)}^2}},
\eea
and the determinant of the metric is
\be\label{det}
\sqrt { - g}  = \sqrt {\frac{{A(r)}}{{B(r)}}} \frac{{{r^2}{\left( {{\Xi ^2}{l^2} - {\Omega ^2}} \right)}}}{{{l^3}}}.
\ee
We find the area of the outer horizon $\mathscr{A}$, by setting $dt = dr = 0$ in the metric (\ref{metric}), and find
\be \label{area}
\mathscr{A} = \int\limits_0^{2\pi } {d\phi } \int\limits_0^L {dz\sqrt { - {g\vert_{dt = dr = 0}}} }  = \frac{{2\pi r_+^2\Xi L}}{l}.
\ee
\subsection{The first law of black hole thermodynamics}
In this subsection, we review the first law of black hole thermodynamics in $f$($T$) gravity. Generally, the first law of black hole thermodynamics
\be
\delta \mathcal{Q}=\tau \delta S, 
\ee
{where $\delta \mathcal{Q}$ and $\delta S$ are the heat flux and the entropy change, respectively,} is violated in $f$($T$) gravity.
The Hawking temperature $\tau=\kappa/2\pi$  in $f$($T$)~gravity, where $\kappa$ is the surface gravity, is the same as one in the Einstein gravity, since it is independent of dynamics of gravity. The black hole solutions in $f$($T$) gravity, violate the Clausius relation $dS = d\mathcal{Q}/\tau$, which suggest that black hols in $f$($T$) gravity, even in a static spacetime, are in nonequilibrium state and produce an intrinsic entropy production \cite{Miao:2011ki}.\\
\indent The heat flux $\delta \mathcal{Q}$ along a Killing vector $\xi^\mu$, is given by
\bea\nonumber
\delta \mathcal{Q} &=& \frac{\kappa }{{2\pi }}\left( {\frac{{f'\left( T \right)d\mathscr{A}}}{4}} \right)\Big| _0^{d\lambda }\\\label{heat} 
&\quad& + \frac{1}{{8\pi }}\int\limits_H {{k^\nu }{\partial ^\mu }f'\left( T \right)\left( {{\xi ^\rho }{S_{\rho \nu \mu }} - {\partial _\nu }{\xi _\mu }} \right)d\mathscr{A}d\lambda },
\eea
where $H$ is the black hole horizon, $\lambda$ is the affine parameter, $k^\mu=dx^\mu/d\lambda$ is the tangent vector to $H$, and $\kappa$ is the surface gravity of the surface $H$. The first term in the right-hand side of Eq. (\ref{heat}) provides the first law of black hole thermodynamics Ref. \cite{Miao:2011ki}. However, the second term, in general, is not equal to zero. This term maybe regarded as a contribution to the intrinsic entropy production $\delta {S_i}$, where
\be
\frac{1}{{8\pi }}\int {{k^\nu }{\partial ^\mu }} f'\left( T \right)\left( {{\xi ^\rho }{S_{\rho \nu \mu }} - {\partial _\nu }{\xi _\mu }} \right)d\mathscr{A}d\lambda  =  - \tau \delta {S_i}.\label{ne}
\ee
The Eq. (\ref{ne}) suggests that the $f$($T$) black holes are in nonequilibrium  thermodynamics, where
\be
\delta \mathcal{Q} = \tau \delta S - \tau \delta {S_i}.
\ee
\indent Miao et al. \cite{Miao:2011ki} showed that the first law of thermodynamics for the $f$($T$) black holes can be recovered  approximatively, if $f''\left(T\right)\ll 1$. In this approximation, the intrinsic entropy production term in Eq. (\ref{heat}) can be neglected, and the entropy of black holes in $f$($T$) gravity becomes
\be \label{entropy1}
S = {\frac{{f'\left( T \right)\mathscr{A}}}{4}}.
\ee
We note that in the case of $f'\left( T \right)=1$, the entropy production $\delta {S_i}$ vanishes, and the entropy (\ref{entropy1}) reduces to the Bekenstein-Hawking entropy in Einstein gravity. 
\begin{figure}[H]
	\centering
	\includegraphics[width=0.8\linewidth]{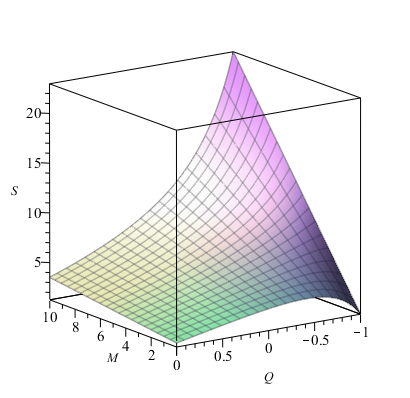}
	\caption{$S$ as a function of $M$ and $Q$. In these plots, we set in which, $\Omega=1,~l=1,~|\alpha|=0.05,~L=1,~\text{and}~r_+=1$.}
	\label{fig:0}
\end{figure}
\begin{widetext}
\noindent Using Eqs. (\ref{ft}) and (\ref{area}) in  (\ref{entropy1}), we find the entropy of {black holes in Eq. (\ref{metric})}, as
\be\label{entropy2}
S = \frac{{\pi \Xi L\left( {7{r_ + }^6 + 9\sqrt {6\left| \alpha  \right|} Q{r_ + }^4 + 18M\left| \alpha  \right|{r_ + }^3 - 54{Q^2}\left| \alpha  \right|{r_ + }^2 - 42\sqrt {6{{\left| \alpha  \right|}^3}} {Q^3}} \right)}}{{9l{{\left( {Q\sqrt {6\left| \alpha  \right|}  + {r_ + }^2} \right)}^2}}}.
\ee
\indent Figure \ref{fig:0} shows the behavior of the entropy (\ref{entropy2}) versus the black hole parameters $M$ and $Q$. Moreover, Figs. \ref{fig:00a} and \ref{fig:00b} show the entropy versus the $\Omega$ and $|\alpha|$, respectively.
\begin{figure}[H]
	\centering
	\begin{subfigure}{0.325\linewidth}
		\includegraphics[width=\linewidth]{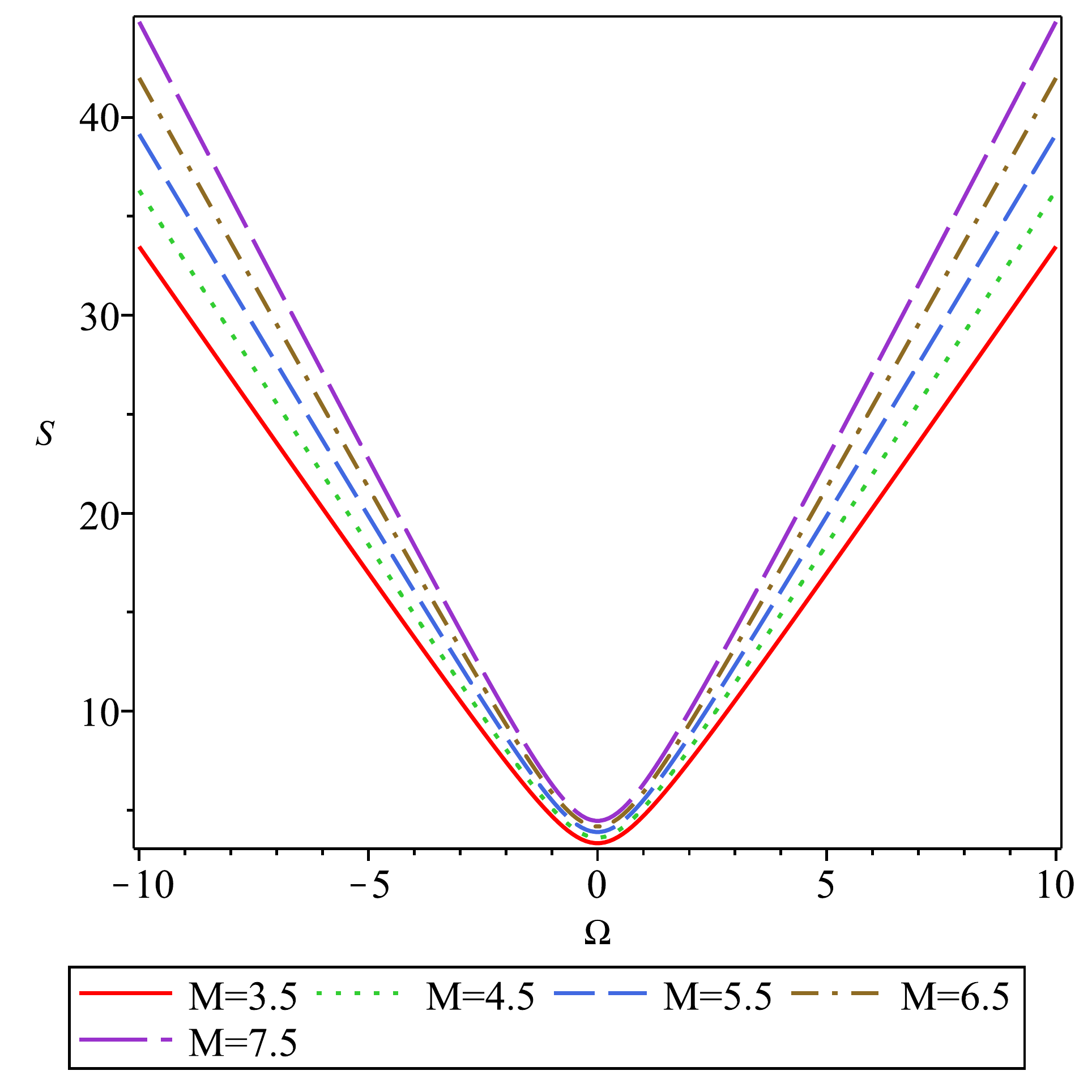}
		\caption{} 
		\label{fig:00a}
	\end{subfigure}
	\begin{subfigure}{0.325\linewidth}
		\includegraphics[width=\linewidth]{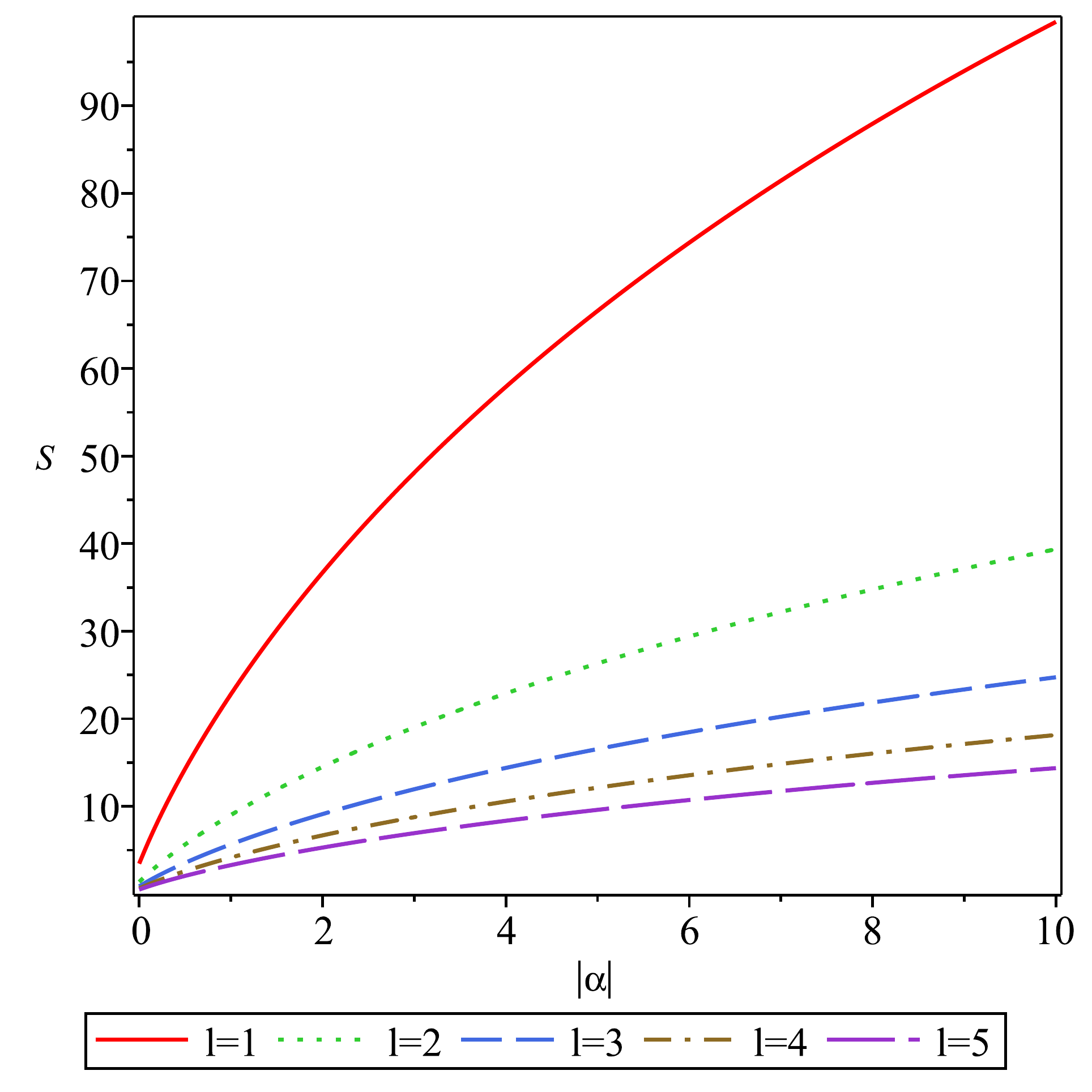}
		\caption{} 
		\label{fig:00b}
	\end{subfigure}
	\caption{(a) $S$ as a function of $\Omega$, with $Q=1$, $l=1$, $\alpha=0.05$, $L=1$, $r_{+}=1$, and several numerical values of $M$. (b) $S$ as a function of $|\alpha|$, with $M=3.5$, $Q=1$, $\Omega=1$ $L=1$, $r_{+}=1$, and several numerical values of $l$. }
	\label{fig:00}
\end{figure}
\end{widetext}
\section{Massless scalar wave equation} 
\indent We consider a massless scalar field $\psi$, in the background of the rotating charged AdS black holes (\ref{metric}), in quadratic $f(T)$ gravity. The scalar wave equation is given by
\cite{Bernard:2017rcw}
\be\label{kge}
\frac{1}{{\sqrt { - g} }}{\partial _\mu }\left( {\sqrt { - g} {g^{\mu \nu }}{\partial _\nu }\psi } \right) = 0.
\ee
We consider the following ansatz for the scalar field 
\be \label{ansatz}
\psi \left( {t,r,z,\phi } \right) = {e^{ - i\omega t + ikz + im\phi }}R\left( r \right),
\ee
where $\omega$ is the frequency of the scalar field, $m$ is the azimutal harmonic index, and $k$ is the wave number. Substituting Eqs. (\ref{contra}), (\ref{det}), and (\ref{ansatz}) into Eq. (\ref{kge}) , we find the radial equation
\begin{widetext}
\be \label{rad1}
B(r) \frac{d^2R\left( r \right)}{dr^2} + \left( {rB(r)\frac{dA(r)}{dr} + rA(r)\frac{dB(r)}{dr} + 4A(r)B(r)} \right)\frac{dR(r)}{dr} + V(r)R\left( r \right) = 0,
\ee
where the potential $V(r)$, is given by
\be
V(r) = \frac{{{r^2}{{\left( {\Xi {l^2}\omega  - \Omega m} \right)}^2} - A(r){l^2}\left\{ {{k^2}{l^4}{\Xi ^4} + {k^2}{\Omega ^4} + {l^2}\left[ {{m^2}{\Xi ^2} - 2m\Xi \Omega \omega  + {\Omega ^2}\left( {{\omega ^2} - 2{\Xi ^2}{k^2}} \right)} \right]} \right\}}}{{A(r){r^2}{\left( {{\Xi ^2}{l^2} - {\Omega ^2}} \right)}^2}}.
\ee
\end{widetext}
In the near-horizon region, we expand the metric function $A(r)$ as a quadratic polynomial in $(r-r_+)$, such as
\be
A(r) \simeq K\left( {r - {r_ + }} \right)\left( {r - {r_ * }} \right),
\ee
where
\be
K = 15{r_ + }^4{\Lambda _{eff}} - 3M{r_ + } + \frac{{3{Q^2}}}{2},
\ee
\be
{r_ * } = {r_ + } - \frac{{2{r_ + }\left( {2{r_ + }^4{\Lambda _{eff}} - M{r_ + } + {Q^2}} \right)}}{{10{r_ + }^4{\Lambda _{eff}} - 2M{r_ + } + {Q^2}}}.
\ee
We note that ${r_{*}}$ is not necessarily any of the black hole horizons. In the near-horizon region, we consider the low-energy limit for the scalar fields, where $r_+ \ll \frac{1}{\omega}$. Moreover, we consider a limit where the outer horizon $r_+$ is very close to ${r_ * }$, in which, $\left| {{r_ + } - {r_ * }} \right| \ll {r_ + }$. Using these two approximations, we find that the radial equation (\ref{rad1}) simplifies to
\begin{widetext}
\be\label{rad2}
\frac{d}{dr}\left\{\left( {r - {r_ + }} \right)\left( {r - {r_ * }} \right)\frac{d}{dr}R\left( r \right)\right\}+ \left[ {\left( {\frac{{{r_ + } - {r_ * }}}{{r - {r_ + }}}} \right)\mathcal{A} + \left( {\frac{{{r_ + } - {r_ * }}}{{r - {r_ * }}}} \right)\mathcal{B} + \mathcal{C}} \right]R\left( r \right) = 0,
\ee
where the constants $\mathcal{A} $, $\mathcal{B}$ and $\mathcal{C} $ are given by
\bea \label{A}
\mathcal{A} &=& \frac{{\mathcal{D}{m^2} + \mathcal{E}m\omega }}{{{K^2}{r_ + }^2{r_ * }^3{\left( {{\Xi ^2}{l^2} - {\Omega ^2}} \right)}^2{{\left( {{r_ + } - {r_ * }} \right)}^2}\beta }} + \frac{{\mathcal{F}{\omega ^2}}}{{K{r_ + }^2{\left( {{\Xi ^2}{l^2} - {\Omega ^2}} \right)}^2{{\left( {{r_ + } - {r_ * }} \right)}^2}\beta }} - C_1,\\\label{B}
\mathcal{B} &=& \frac{{\mathcal{G}{m^2} + \mathcal{I}m\omega }}{{{K^2}{r_ + }^3{r_ * }{\left( {{\Xi ^2}{l^2} - {\Omega ^2}} \right)}^2{{\left( {{r_ + } - {r_ * }} \right)}^2}\beta }} + \frac{{\mathcal{J}{\omega ^2}}}{{K{r_ + }^2{\left( {{\Xi ^2}{l^2} - {\Omega ^2}} \right)}^2{{\left( {{r_ + } - {r_ * }} \right)}^2}\beta }} + C_2,
\eea
\end{widetext}
\be \label{C}
\mathcal{C} =  - \frac{{2m\Omega {{\left( {\Xi {l^2}\omega  - \Omega m/2} \right)}^2}\left( {{r_ + }^2 + {r_ + }{r_ * } + {r_ * }^2} \right)}}{{{\left( {{\Xi ^2}{l^2} - {\Omega ^2}} \right)}^2{K^2}{r_ + }^3{r_ * }^3\beta }}.
\ee
In Eqs. (\ref{A})--(\ref{B}), the constants $C_1$ and $C_2$ are given by $C_1=C_2={{k^2}{l^2}}/{K\beta {{\left( {{r_ + } - {r_ * }} \right)}^2}{r_ + }^2}$, and
\be
\mathcal{D} = {\Omega ^2}\left( {{r_ + }^3 + 2{r_ + }^2{r_ * } + 3{r_ * }^2{r_ + }} \right) - {l^4}{\Xi ^2}{r_ * }^3K,
\ee
\be
\mathcal{E} = 2\Omega \Xi {l^2}\left( {K{l^2}{r_ * }^3 - {r_ + }^3 - 2{r_ + }^2{r_ * } - 3{r_ + }{r_ * }^2} \right),
\ee
\be
\mathcal{F} =  - {l^4}{\Omega ^2},
\ee
\be
\mathcal{G} = K{\Xi ^2}{l^4}{r_ + }{r_ * } - {\Omega ^2}\left( {3{r_ + }^2 + 2{r_ + }{r_ * } + {r_ * }^2} \right),
\ee
\be
\mathcal{I} = 2{l^2}\Xi \Omega \left[ {3{r_ + }^2 - {r_ + }{r_ * }\left( {K{l^2} - 2} \right) + {r_ * }^2} \right],
\ee
\be
\mathcal{J} = {l^4}{\Omega ^2}.
\ee
\section{Hidden conformal symmetry}
To find the existence of the possible hidden symmetry, we introduce the following conformal coordinates $\omega^+$, $\omega^-$ and $y$, in terms of the black hole coordinates $t,\,r$ and $\phi$ 
\be
\omega^+ = \sqrt{\frac{r-r_+}{r-{r_ * }}}e^{2\pi T_R\phi+2n_R t},
\ee
\be
\omega^- = \sqrt{\frac{r-r_+}{r-{r_ * }}}e^{2\pi T_L\phi+2n_L t},
\ee
\be
y = \sqrt{\frac{r_+-{r_ * }}{r-{r_ * }}}e^{\pi (T_L+T_R)\phi+(n_L+n_R)t},
\ee
where $T_L$, $T_R$, $n_L$ and $n_R$ are constants.
We also define the sets of \textit{local} vector fields
\be \label{h1}
H_1 = i\p_+,
\ee
\be \label{vecfield1}
H_0 = i\left(\omega^+\p_++\frac{1}{2}y\p_y\right),
\ee
\be
H_{-1} = i(\omega^{+2}\p_++\omega^+y\p_y-y^2\p_-),
\ee
as well as 
\be
\bar H_1=i\p_-,
\ee
\be \label{vecfield2}
\bar H_0=i\left(\omega^-\p_-+\frac{1}{2}y\p_y\right),
\ee
\be \label{hbm1}
\bar H_{-1}=i(\omega^{-2}\p_-+\omega^-y\p_y-y^2\p_+).
\ee
The vector fields (\ref{h1})--(\ref{hbm1}) obey the \slr~algebra, as
\be
[H_0,H_{\pm 1}]=\mp i H_{\pm 1}, \quad [H_{-1},H_{1}]=-2iH_0,\label{Hs}
\ee
\be
[\bar H_0,\bar H_{\pm 1}]=\mp i \bar H_{\pm 1}, \quad [\bar H_{-1},\bar H_{1}]=-2i\bar H_0.
\ee
The quadratic Casimir operators of  \slr~algebra, are given by
\begin{widetext}
\be \label{casimir1}
{\cal H}^2=\bar { \cal H}^2=-H_0^2+{\frac{1}{2}}(H_1H_{-1}+H_{-1}H_1) = {\frac{1}{4}}(y^2\p_y^2-y\p_y )+y^2\p_+\p_-.
\ee
We notice that the Casimir operators (\ref{casimir1}), can be rewritten in terms of $\left(t,r,\phi\right)$ coordinates, as
\bea\nonumber
\mathcal{H} ^2 &=& (r-r_+)(r-{r_ * })\frac{\p^2}{\p r^2}+(2r-r_+-{r_ * })\frac{\p}{\p r} + \left(\frac{r_+-{r_ * }}{r-{r_ * }}\right)\left[\left(\frac{n_L-n_R}{4\pi G}\p_\phi -\frac{T_L-T_R}{4G}\p_t\right)^2+ C_2\right] \\\label{casimir2}
&\quad& -\left(\frac{r_+-{r_ * }}{r-{r_ + }}\right)\left[\left(\frac{n_L+n_R}{4\pi G}\p_\phi -\frac{T_L+T_R}{4G}\p_t\right)^2- C_1\right],
\eea
where $G=n_LT_R-n_R T_L$.\\
\indent The Casimir operator (\ref{casimir2}) reproduces the radial equation (\ref{rad2}), by choosing the right and left temperatures, as 
\be \label{tempR}
{T_R} = \frac{{{r_ + }K\left( {{r_ + } - {r_ * }} \right)\left( {{\Xi ^2}{l^2} - {\Omega ^2}} \right)\sqrt {\beta {r_ + }{r_ * }\delta } }}{{4\pi \delta }},
\ee
\be \label{tempL}
{T_L} = \frac{{{r_ + }K\left( {{\Xi ^2}{l^2} - {\Omega ^2}} \right)\left[ {{r_ + }^4 + 2{r_ + }^3{r_ * } + 6{r_ + }^2{r_ * }^2 - 2{r_ * }^3{r_ + }\left( {K{l^2} - 1} \right) + {r_ * }^4} \right]\sqrt {\beta {r_ + }{r_ * }\delta } }}{{4\pi {{\left( {{r_ + } + {r_ * }} \right)}^3}\delta }},
\ee
\begin{figure}[H]
	\centering
	\begin{subfigure}{0.3\linewidth}
		\includegraphics[width=\linewidth]{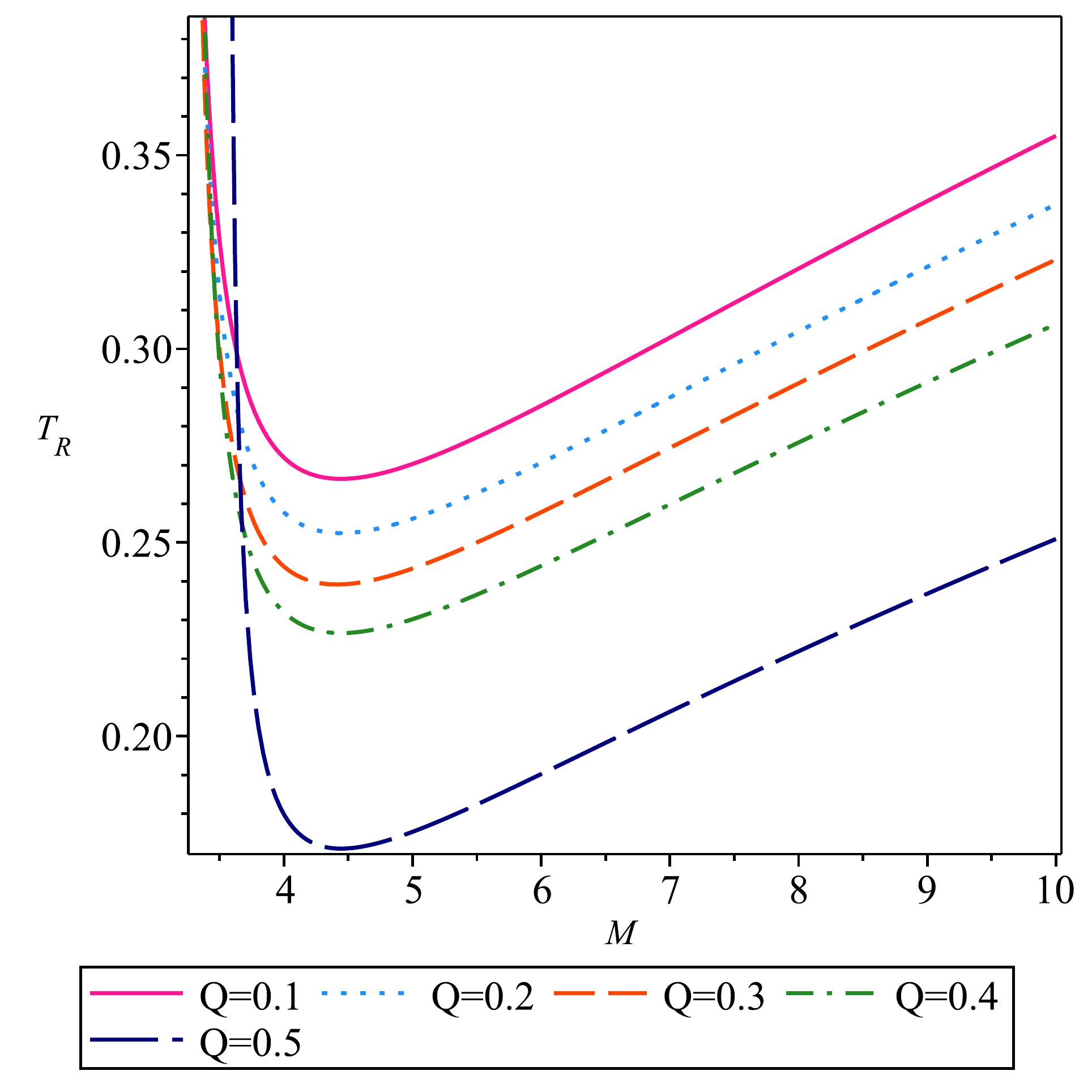} 
		\caption{}
		\label{fig:1a}
	\end{subfigure}
	\begin{subfigure}{0.3\linewidth}
		\includegraphics[width=\linewidth]{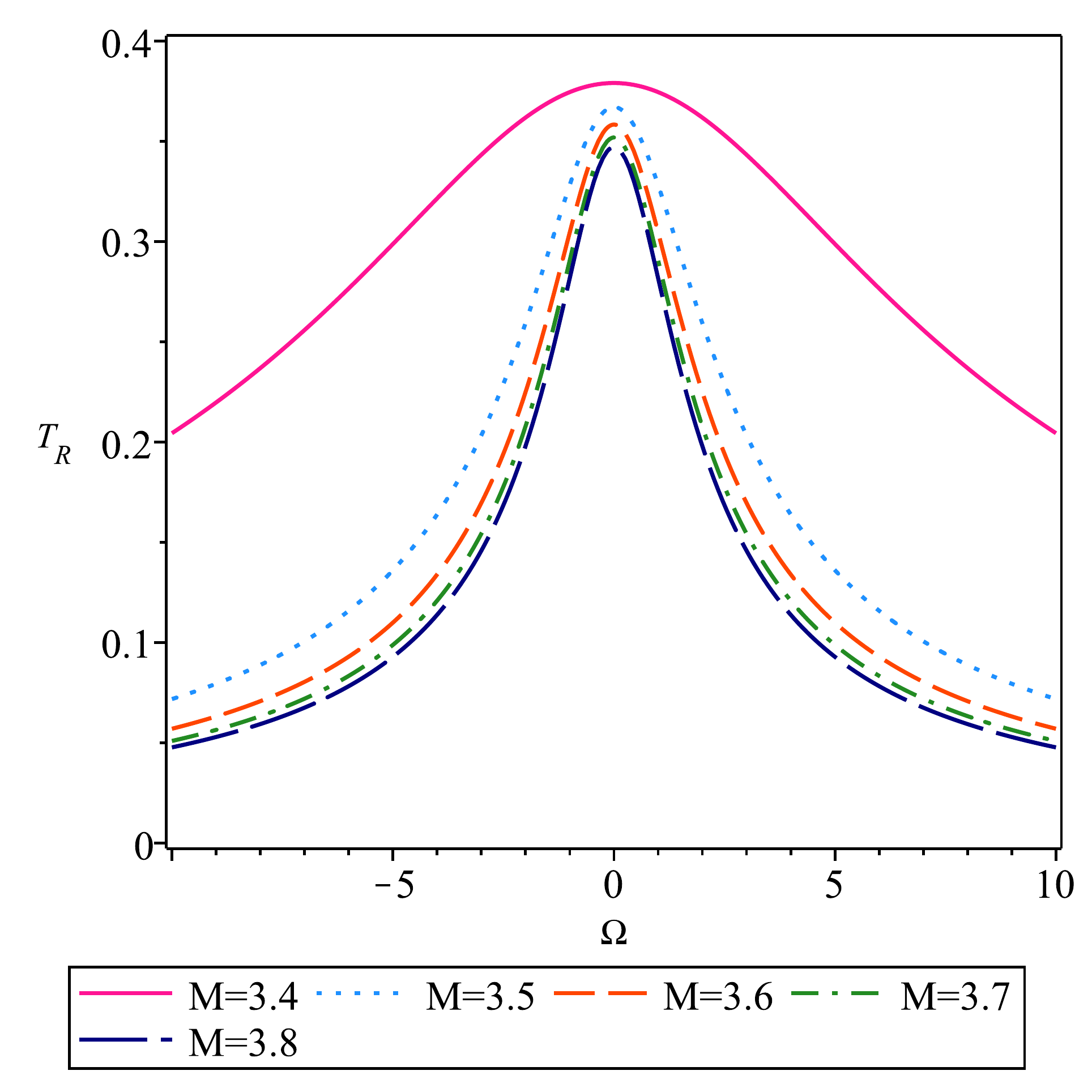}
		\caption{}
		\label{fig:1b}
	\end{subfigure}
	\begin{subfigure}{0.3\linewidth}
		\includegraphics[width=\linewidth]{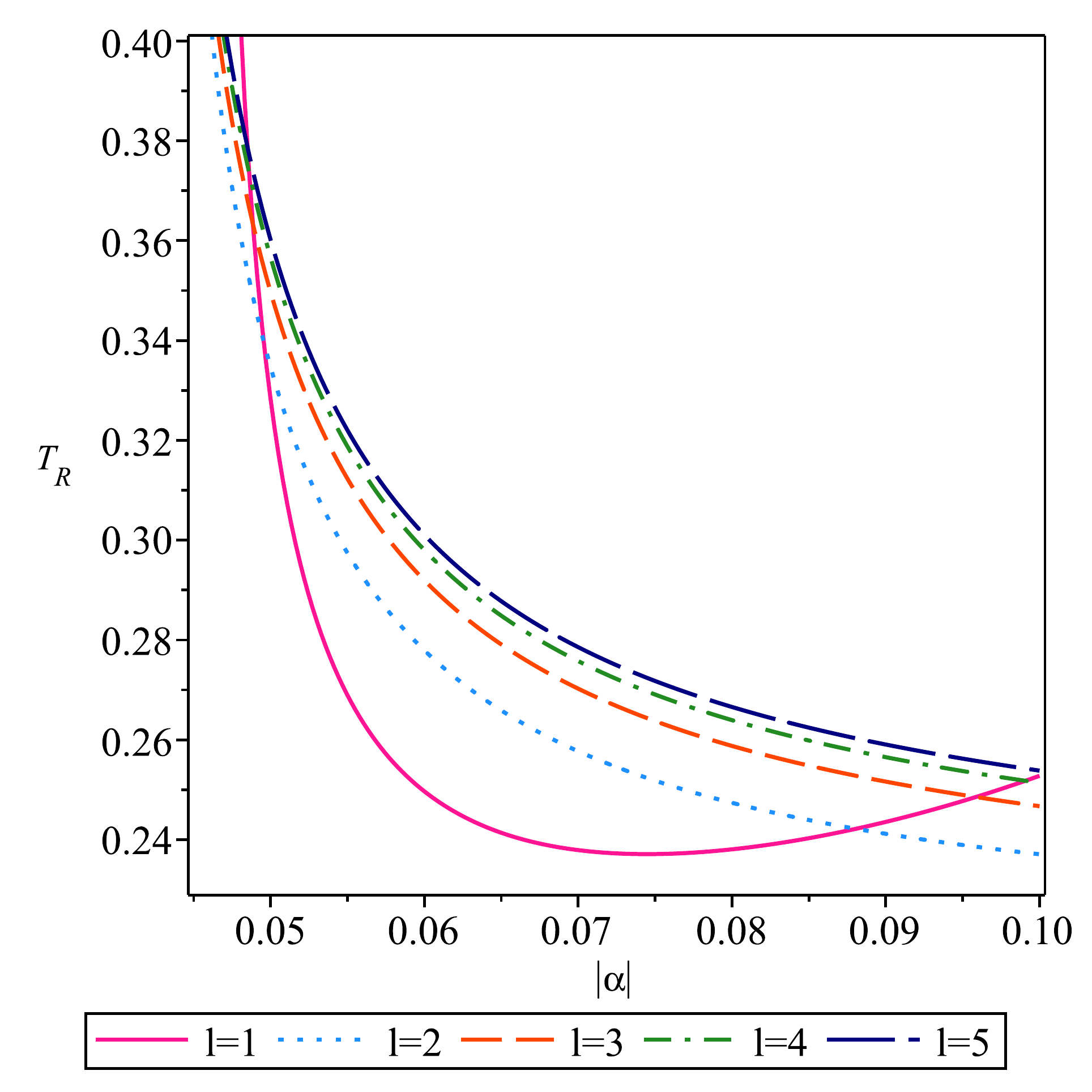}
		\caption{}
		\label{fig:1c}
	\end{subfigure}
	\begin{subfigure}{0.3\linewidth}
		\includegraphics[width=\linewidth]{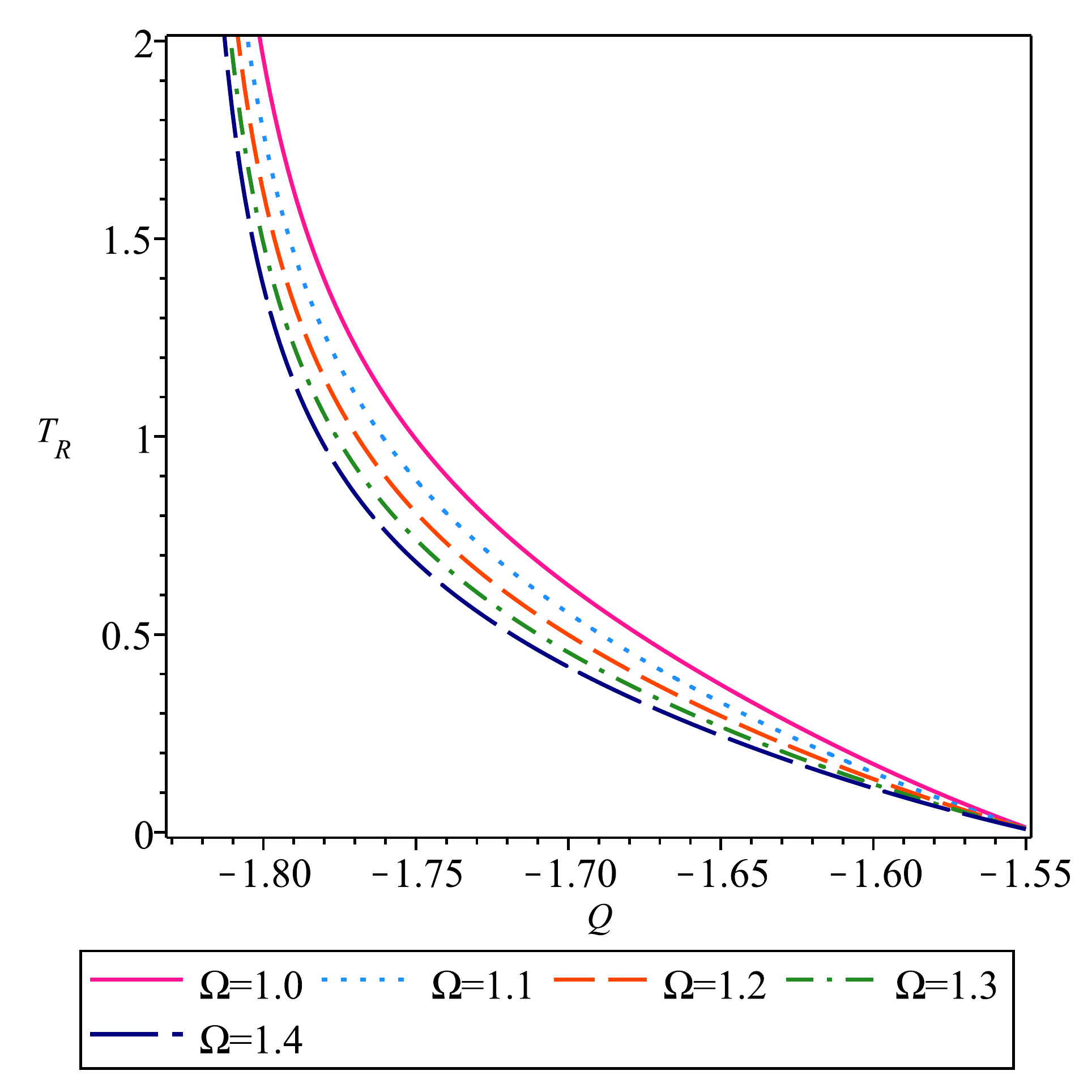}
		\caption{} 
		\label{fig:1d}
	\end{subfigure}
	\begin{subfigure}{0.3\linewidth}
		\includegraphics[width=\linewidth]{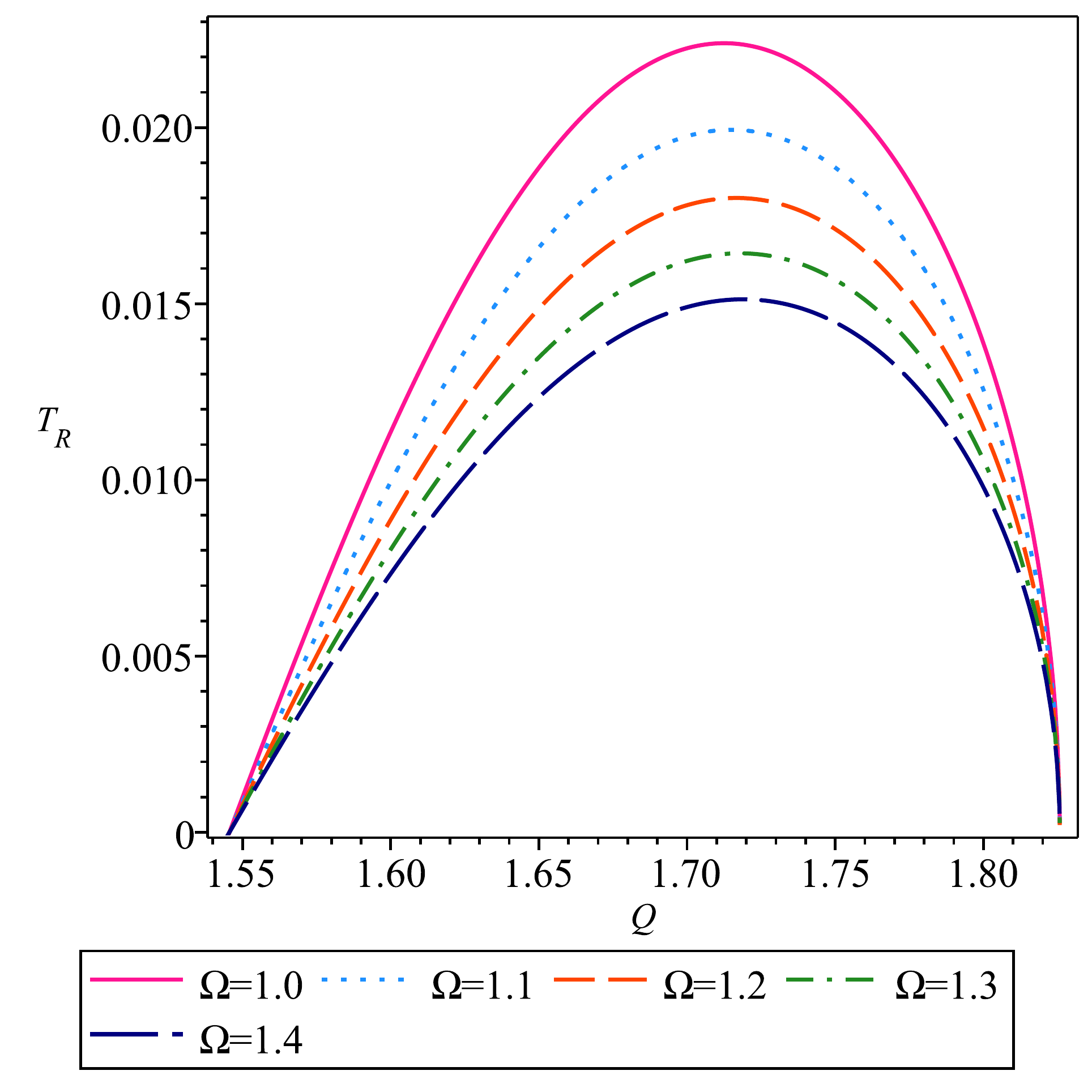}
		\caption{}
		\label{fig:1e}
	\end{subfigure}
	\begin{subfigure}{0.3\linewidth}
		\includegraphics[width=\linewidth]{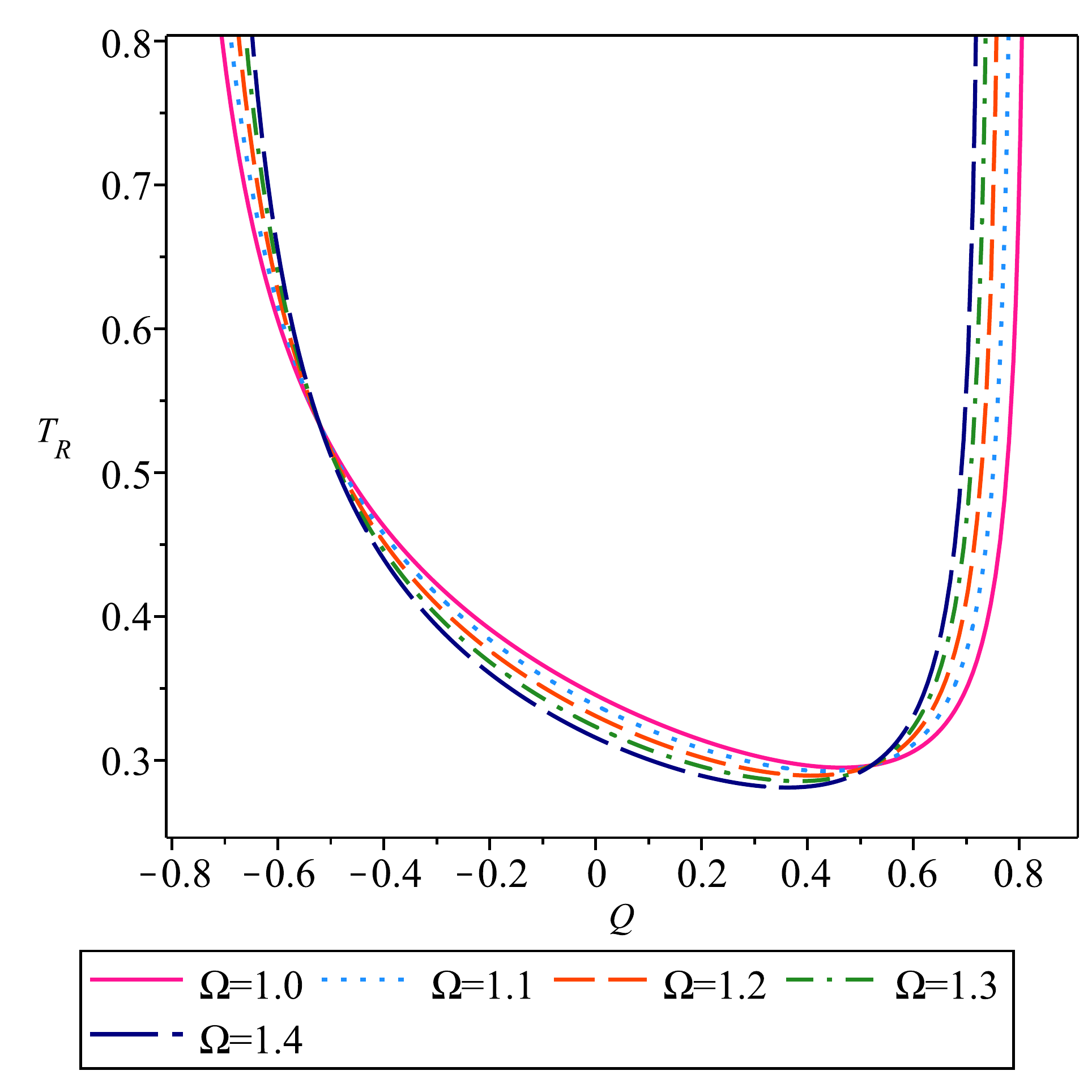}
		\caption{}
		\label{fig:1f}
	\end{subfigure}
	\caption{(a) $T_R$ as a function of $M$, with $\Omega=1,~l=1,~|\alpha|=0.05,~r_+=1$, and several numerical values of $Q$. (b) $T_R$ as a function of $\Omega$, with $Q=0.1,~l=1,~|\alpha|=0.05,~r_+=1$, and several numerical values of $M$. (c) $T_R$ as a function of $|\alpha|$, with $Q=0.1,~M=3.5,~\Omega=1,~r_+=1$, and several numerical values of $l$. (d), (e), and (f) $T_R$ as a function of $Q$, with $M=3.5,~l=1,~|\alpha|=0.05,~r_+=1$, and several numerical values of $\Omega$. }
	\label{fig:1}
\end{figure}
\begin{figure}[H]
	\centering
	\begin{subfigure}{0.3\linewidth}
		\includegraphics[width=\linewidth]{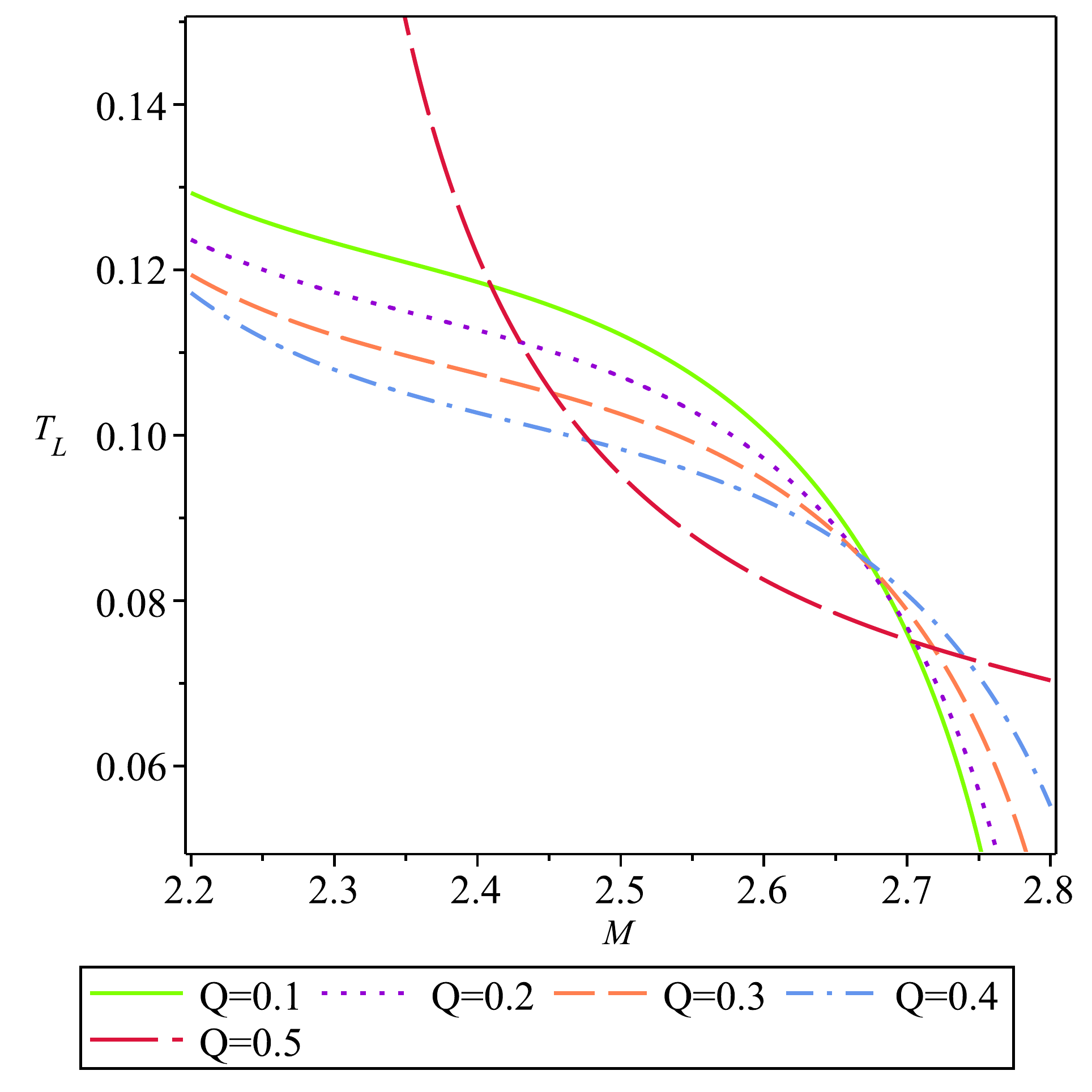}
		\caption{} 
		\label{fig:2a}
	\end{subfigure}
	\begin{subfigure}{0.3\linewidth}
		\includegraphics[width=\linewidth]{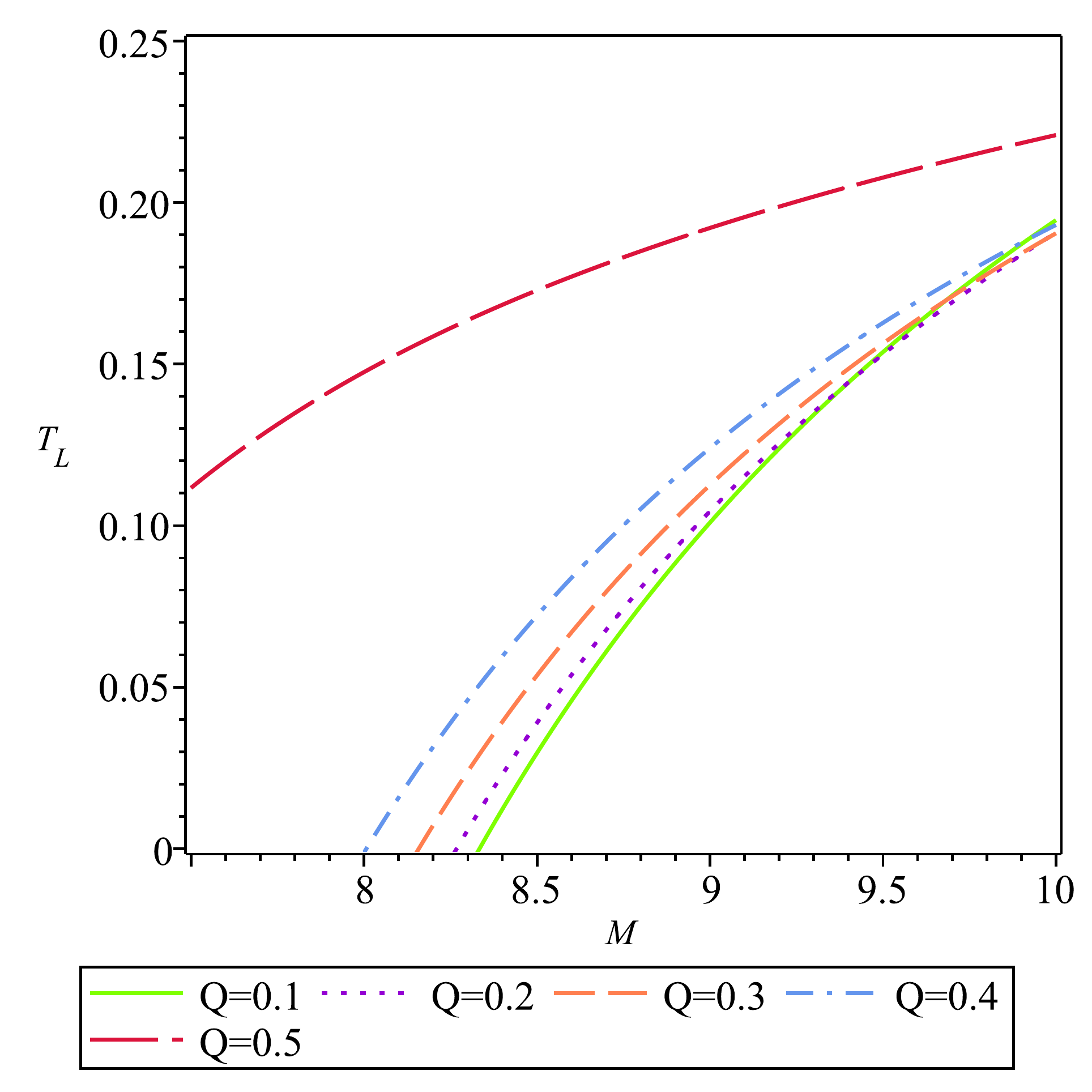}
		\caption{}
		\label{fig:2b}
	\end{subfigure}
	\begin{subfigure}{0.3\linewidth}
		\includegraphics[width=\linewidth]{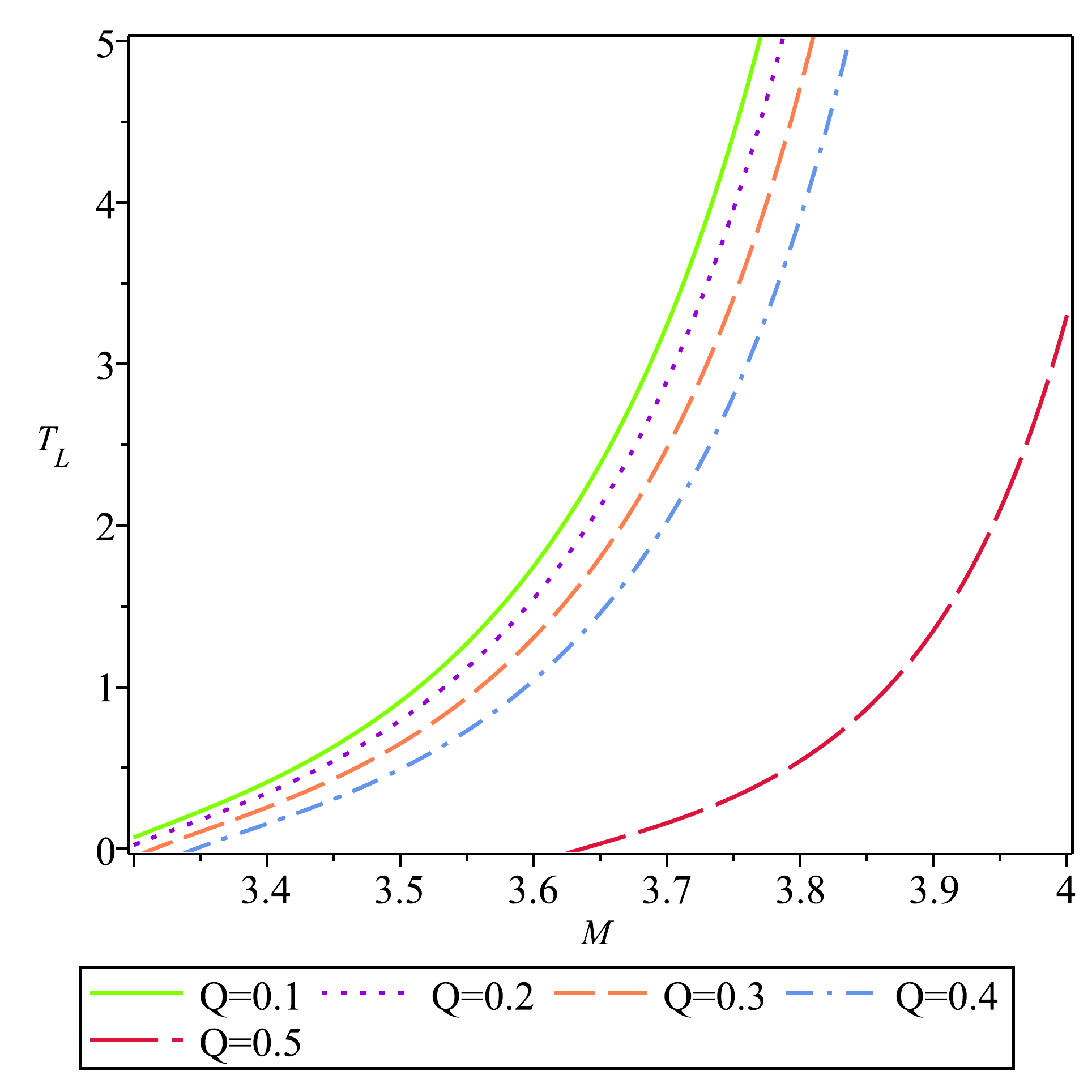}
		\caption{}
		\label{fig:2c}
	\end{subfigure}
	\begin{subfigure}{0.3\linewidth}
		\includegraphics[width=\linewidth]{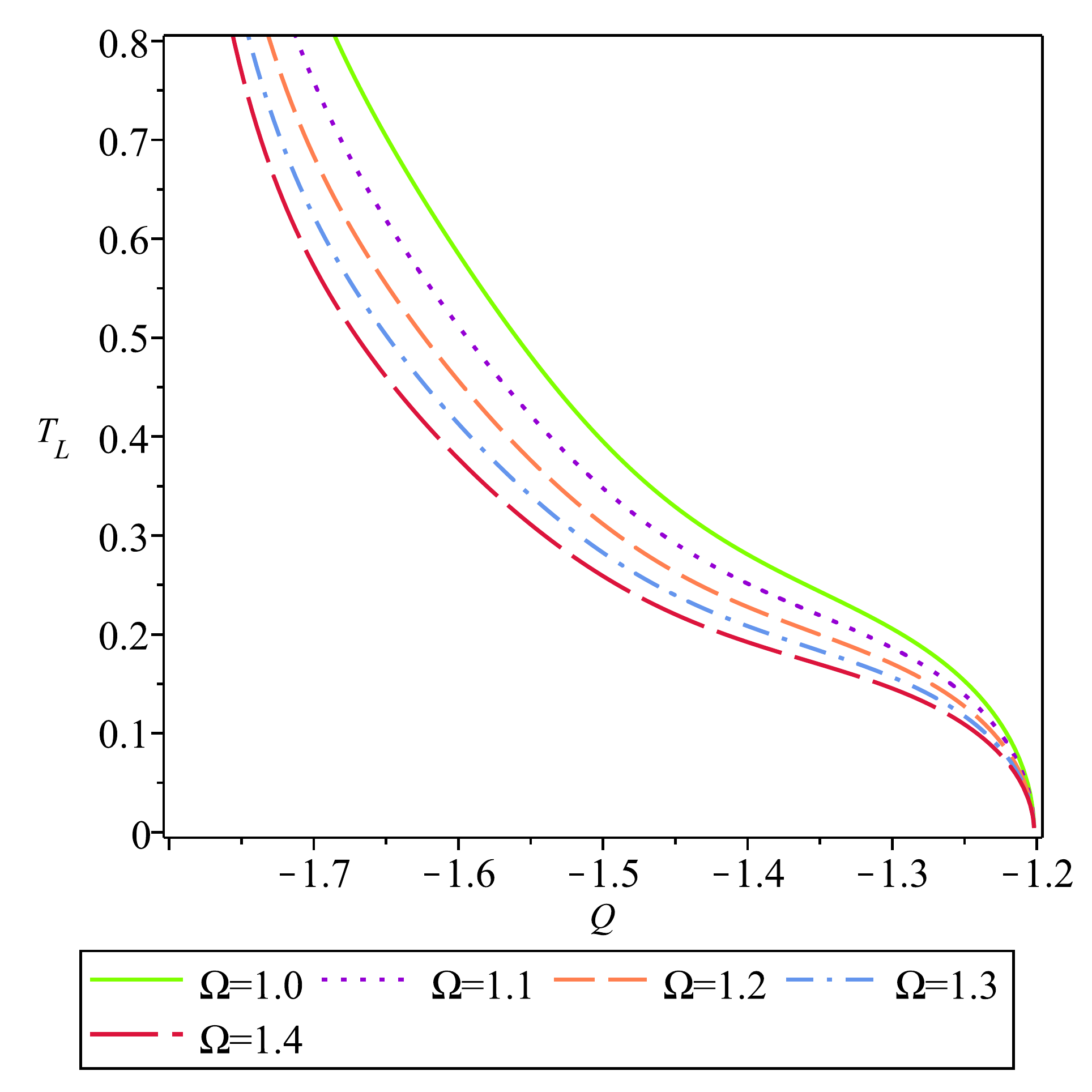}
		\caption{} 
		\label{fig:2d}
	\end{subfigure}
	\begin{subfigure}{0.3\linewidth}
		\includegraphics[width=\linewidth]{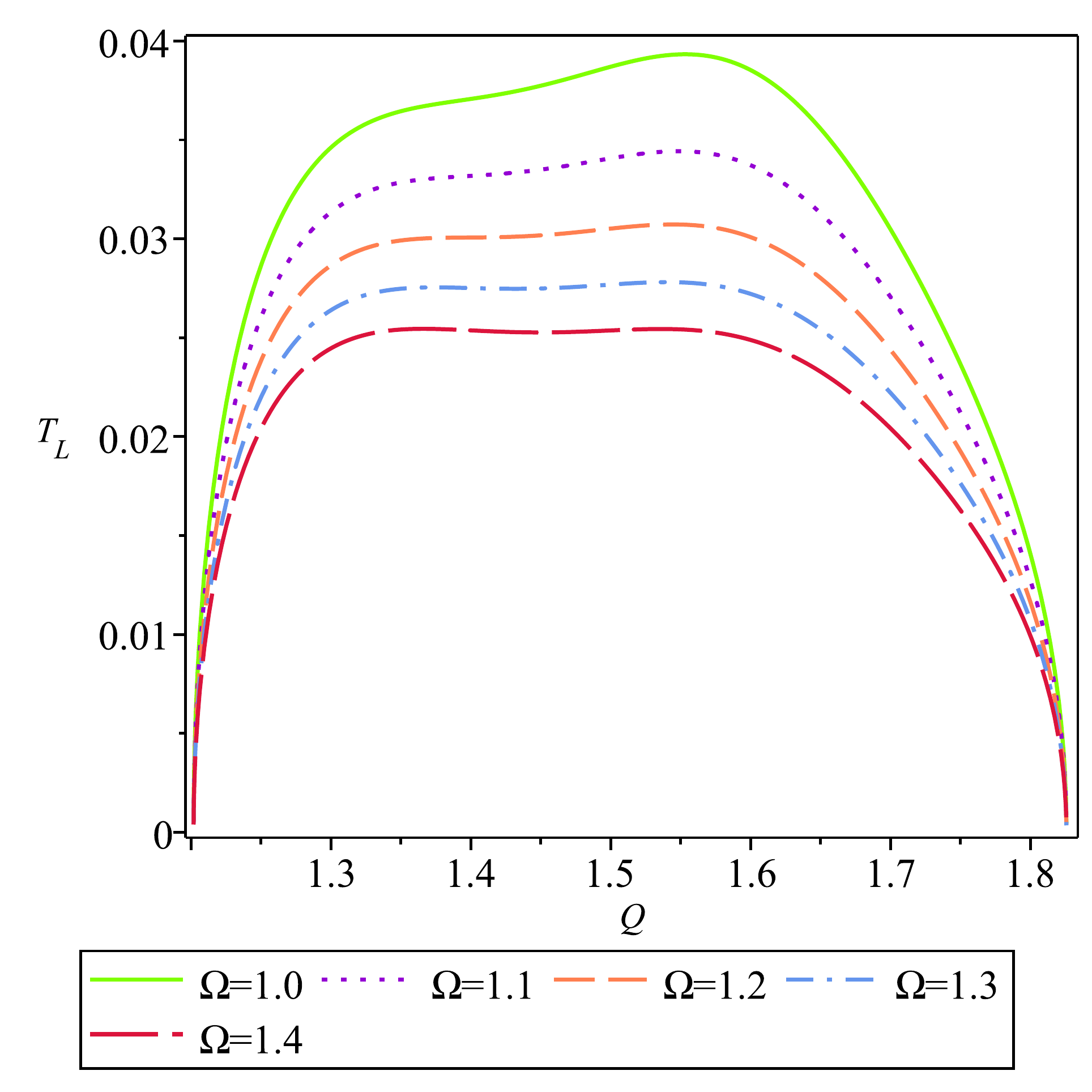}
		\caption{}
		\label{fig:2e}
	\end{subfigure}
	\begin{subfigure}{0.3\linewidth}
		\includegraphics[width=\linewidth]{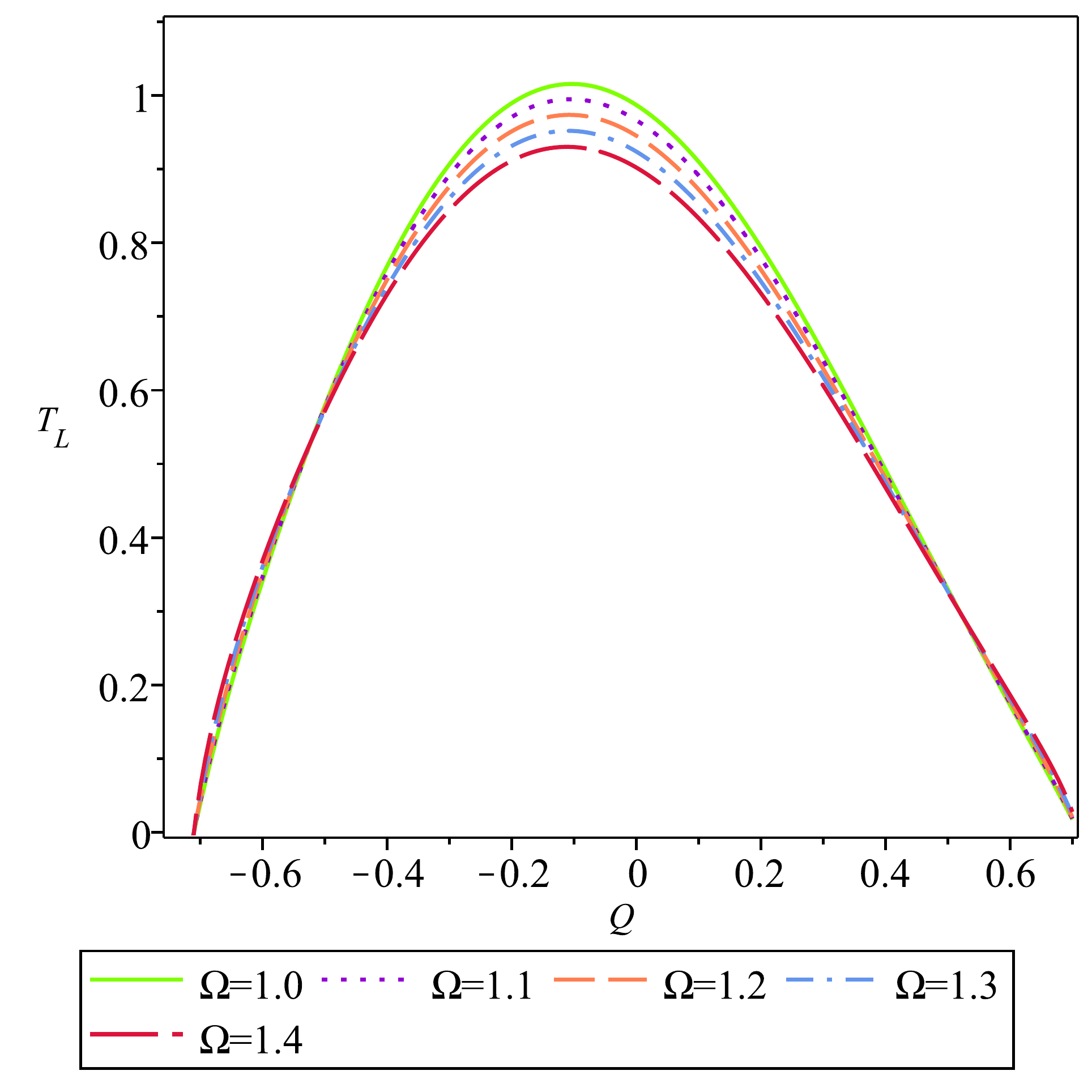}
		\caption{}
		\label{fig:2f}
	\end{subfigure}
	\begin{subfigure}{0.3\linewidth}
		\includegraphics[width=\linewidth]{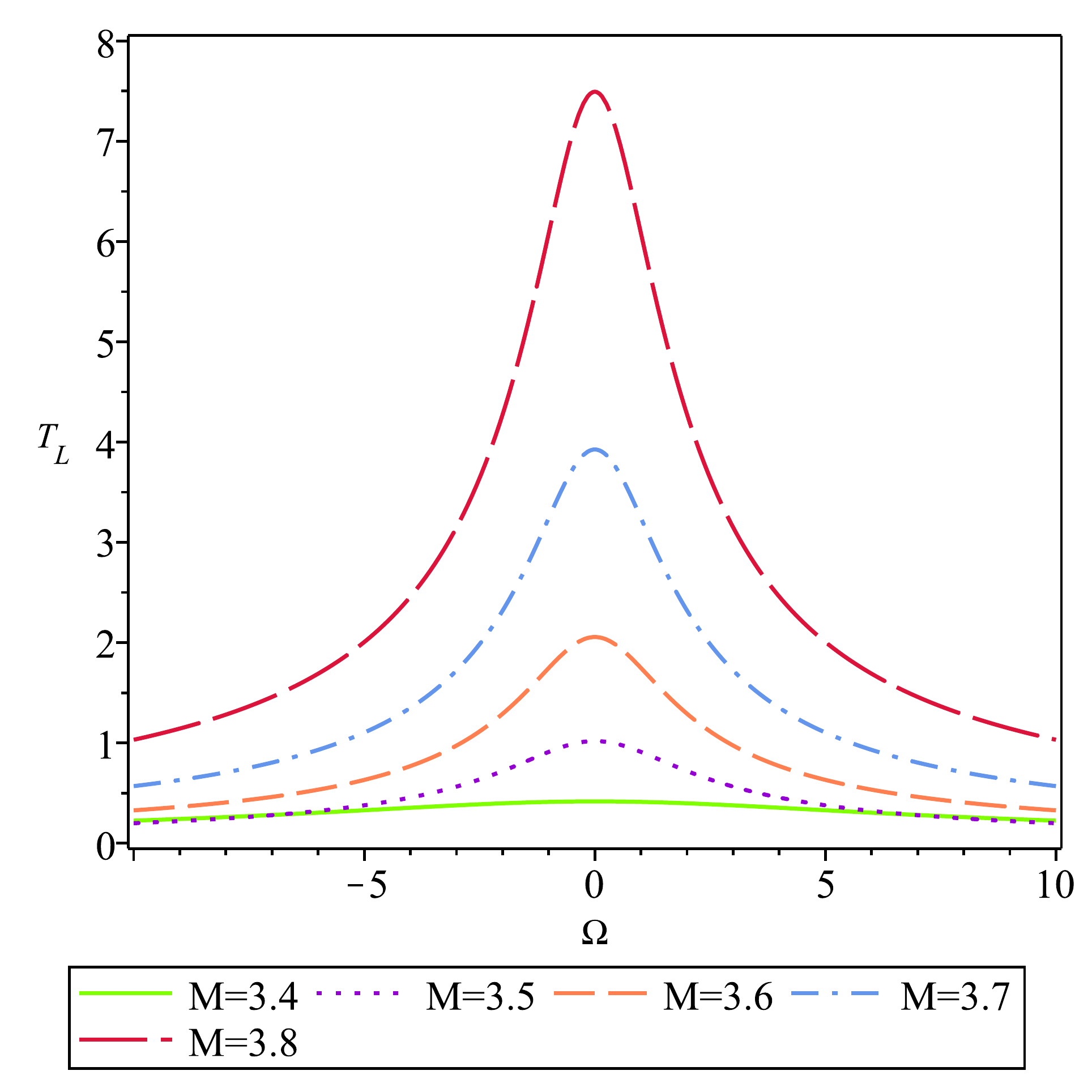}
		\caption{} 
		\label{fig:2g}
	\end{subfigure}
	\begin{subfigure}{0.3\linewidth}
		\includegraphics[width=\linewidth]{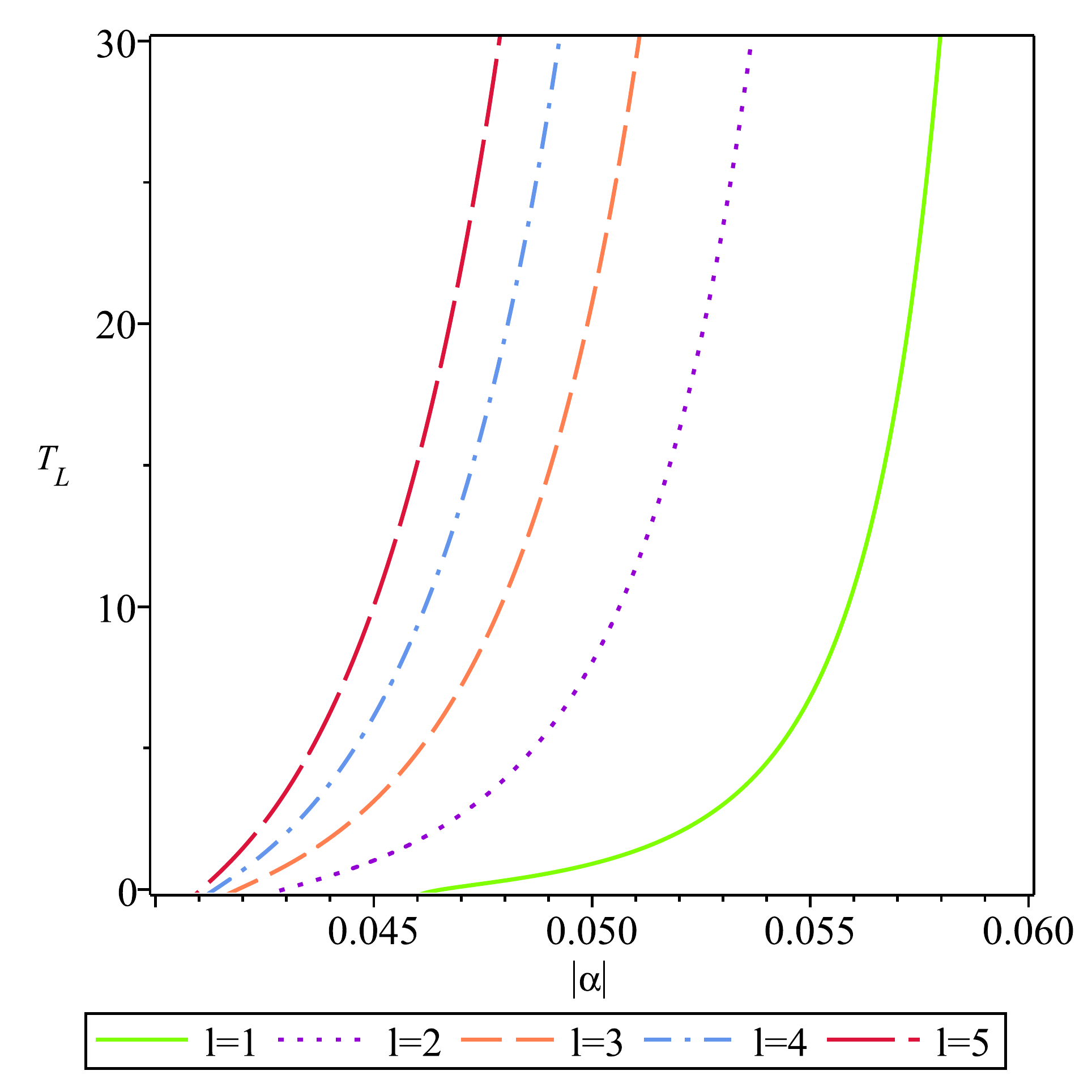}
		\caption{}
		\label{fig:2i}
	\end{subfigure}
	\caption{(a), (b), and (c) $T_L$ as a function of $M$, with $\Omega=1,~l=1,~|\alpha|=0.05,~r_+=1$, and several numerical values of $Q$. (d), (e), and (f) $T_L$ as a function of $Q$, with $M=3.5,~l=1,~|\alpha|=0.05,~r_+=1$, and several numerical values of $\Omega$. (g) $T_L$ as a function of $\Omega$, with $Q=0.1,~l=1,~|\alpha|=0.05,~r_+=1$, and several numerical values of $M$. (h) $T_L$ as a function of $|\alpha|$, with $Q=0.1,~\Omega=1,~M=3.5,~r_+=1$, and several numerical values of $l$.  }
	\label{fig:2}
\end{figure}
\begin{figure}[H]
	\centering
	\begin{subfigure}{0.3\linewidth}
		\includegraphics[width=\linewidth]{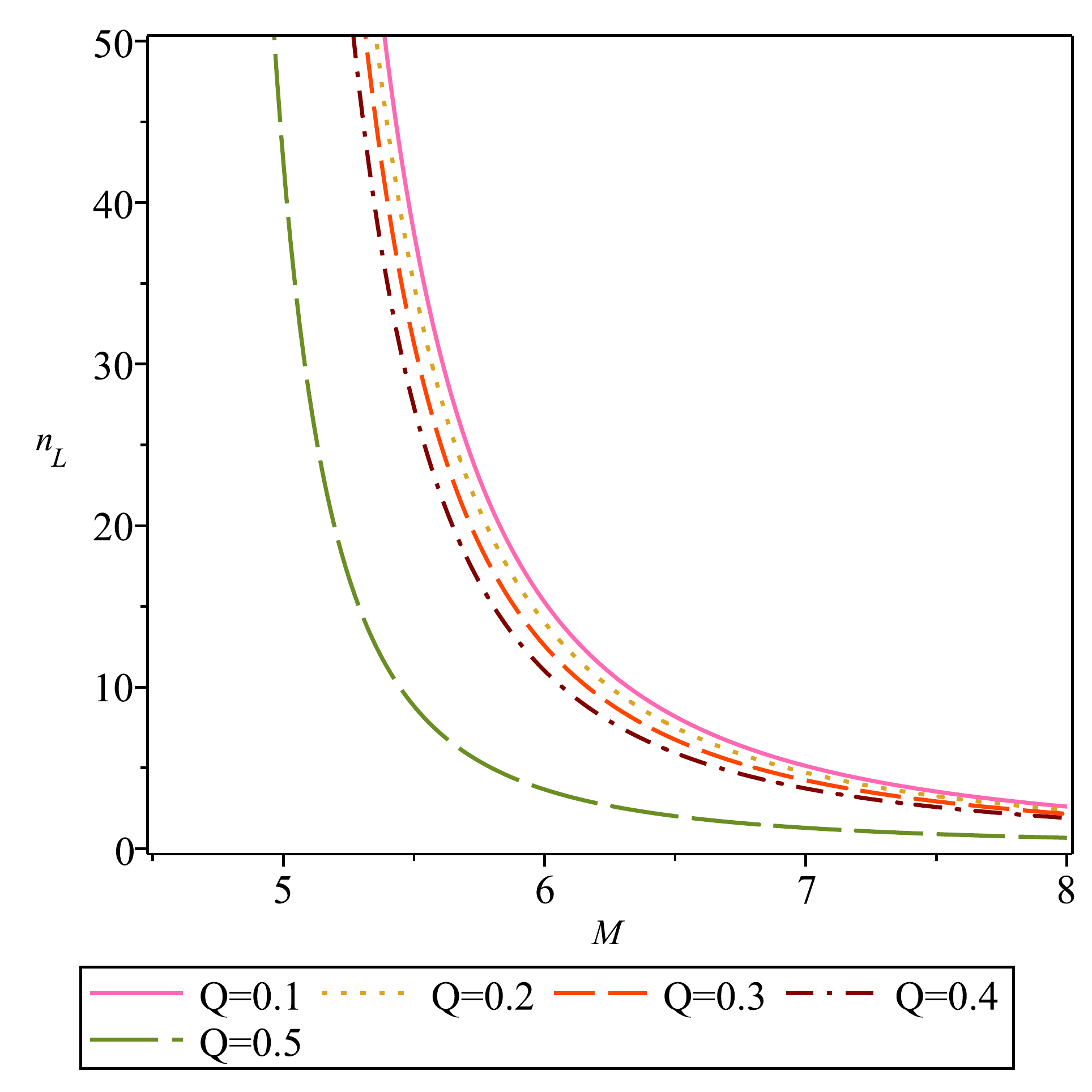}
		\caption{} 
		\label{fig:3a}
	\end{subfigure}
	\begin{subfigure}{0.3\linewidth}
		\includegraphics[width=\linewidth]{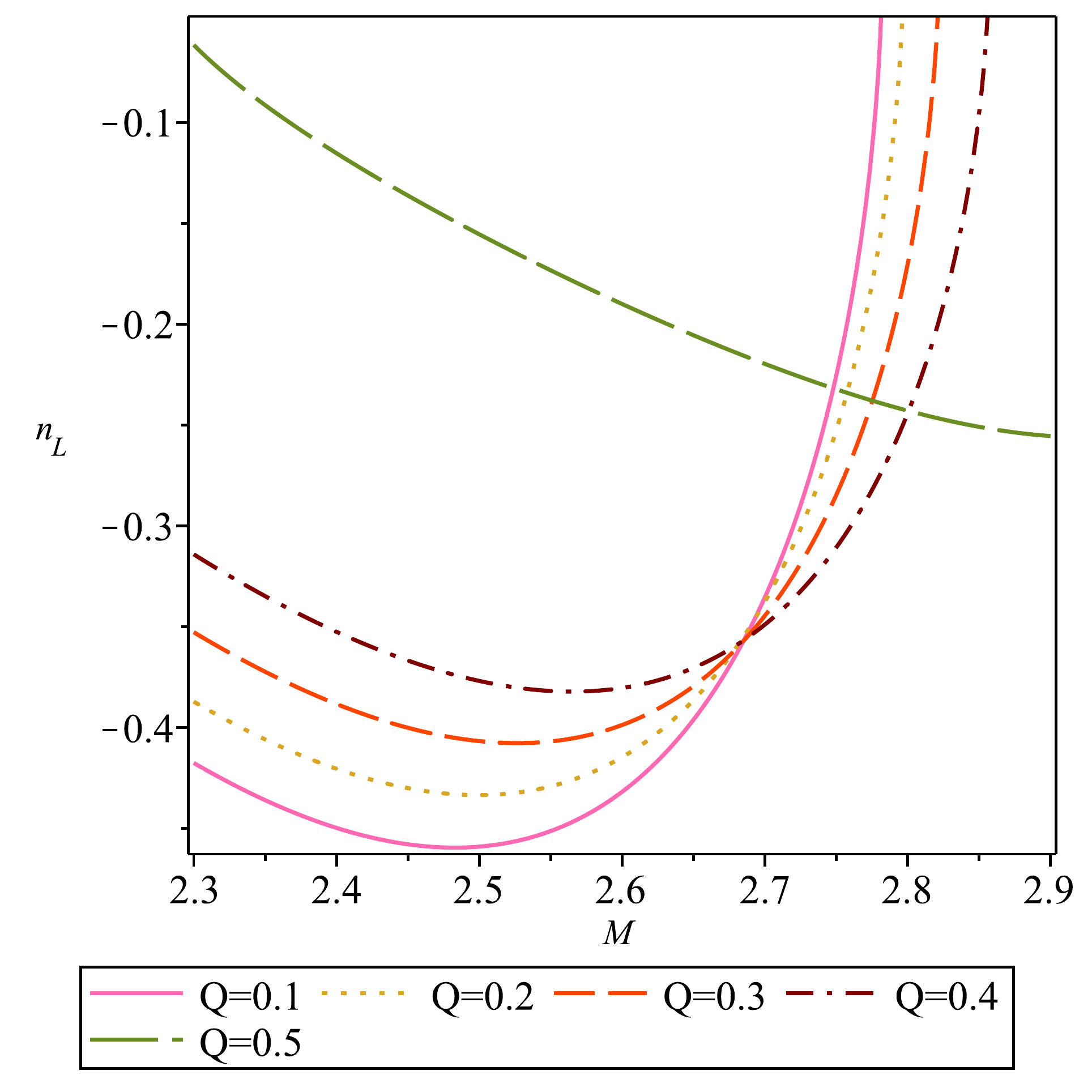}
		\caption{}
		\label{fig:3b}
	\end{subfigure}
	\begin{subfigure}{0.3\linewidth}
		\includegraphics[width=\linewidth]{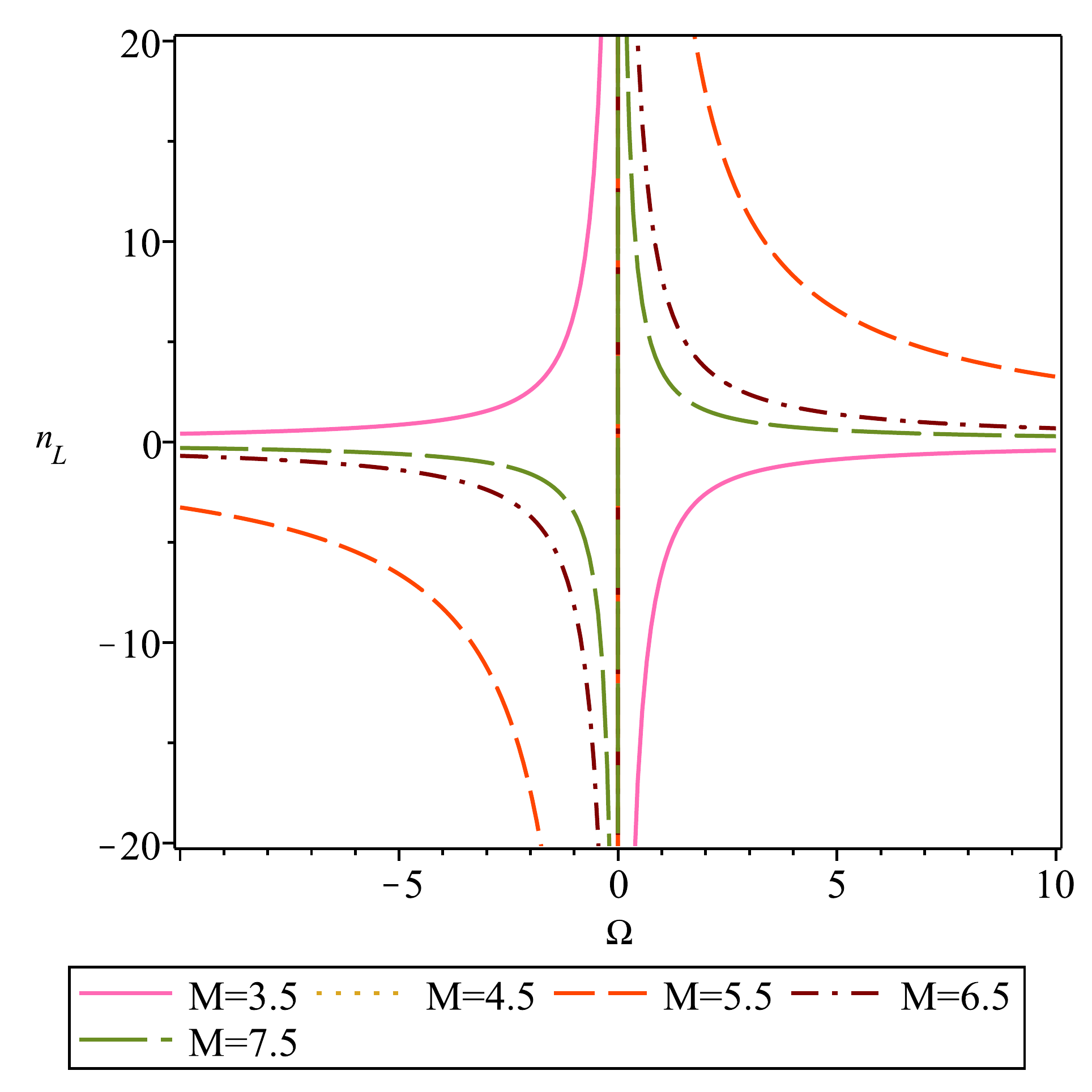}
		\caption{}
		\label{fig:3c}
	\end{subfigure}
	\begin{subfigure}{0.3\linewidth}
		\includegraphics[width=\linewidth]{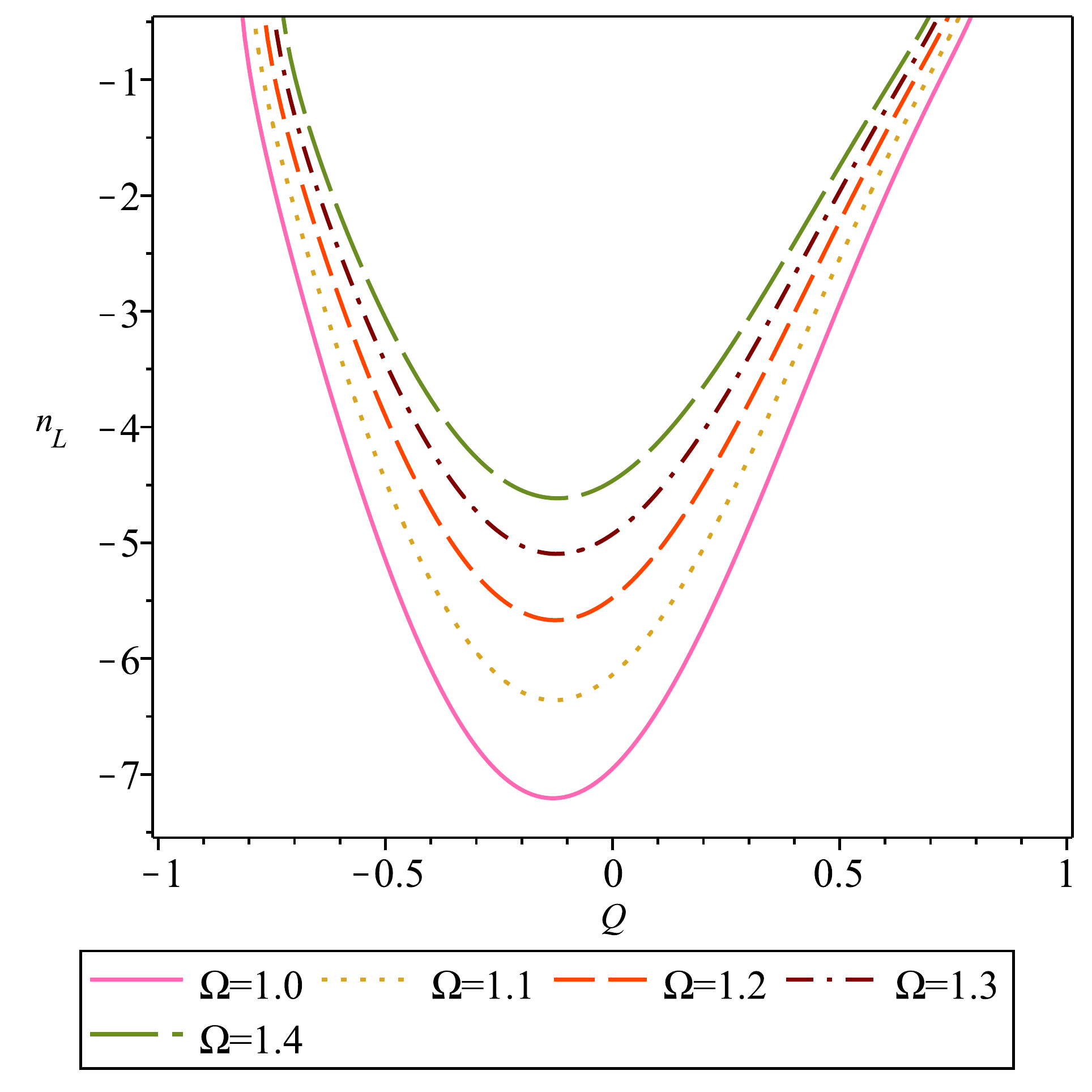}
		\caption{} 
		\label{fig:3d}
	\end{subfigure}
	\begin{subfigure}{0.3\linewidth}
		\includegraphics[width=\linewidth]{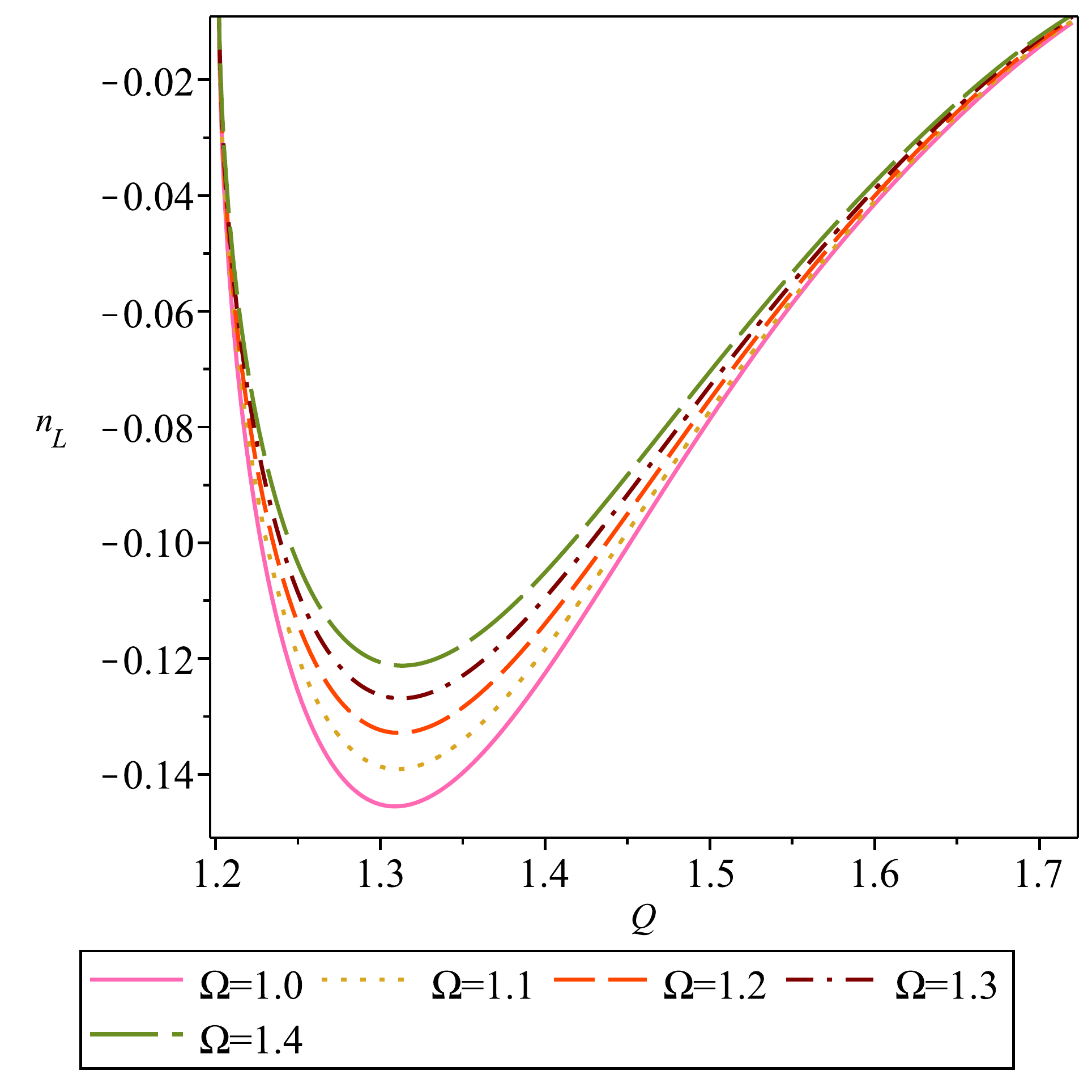}
		\caption{}
		\label{fig:3e}
	\end{subfigure}
	\begin{subfigure}{0.3\linewidth}
		\includegraphics[width=\linewidth]{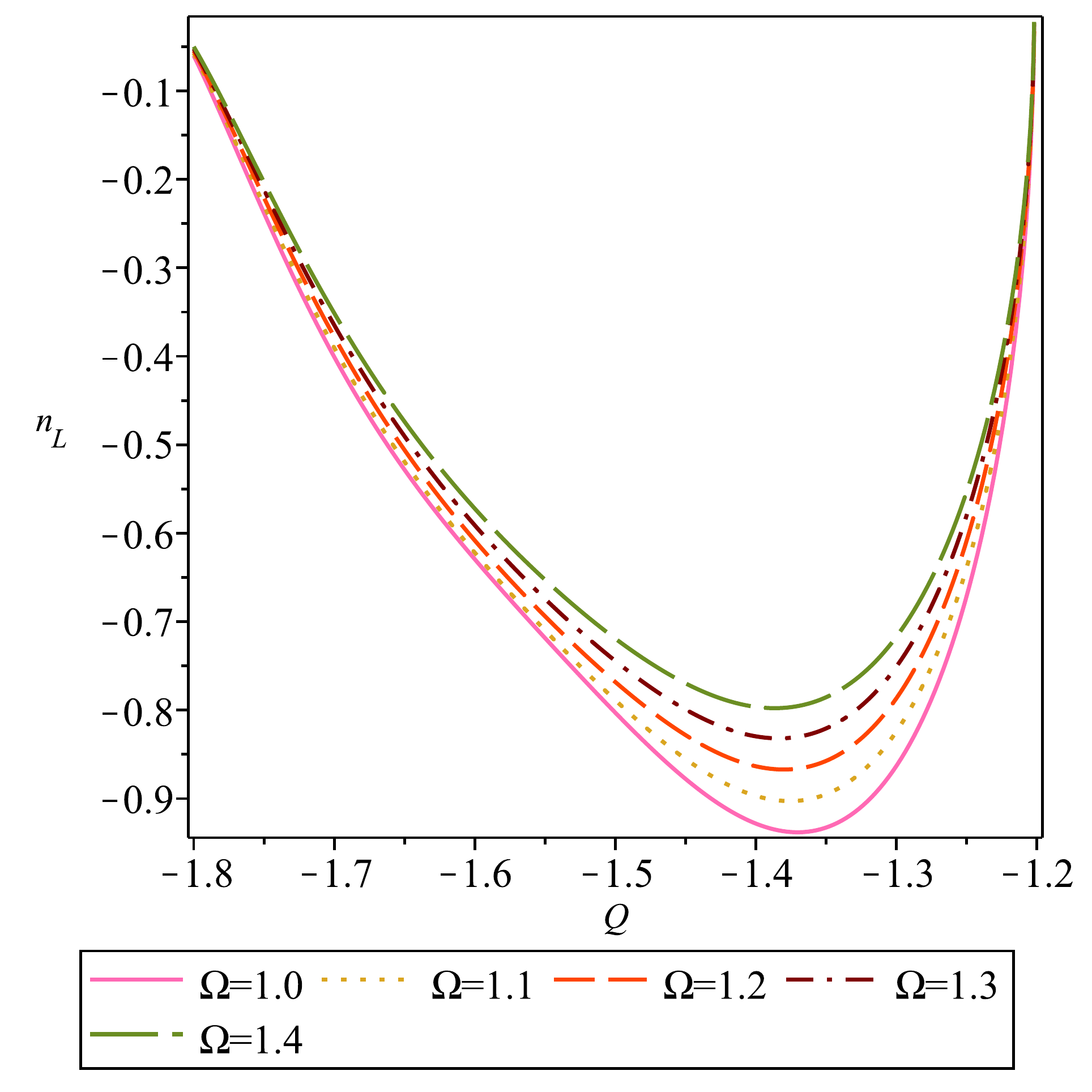}
		\caption{}
		\label{fig:3f}
	\end{subfigure}
	\begin{subfigure}{0.3\linewidth}
		\includegraphics[width=\linewidth]{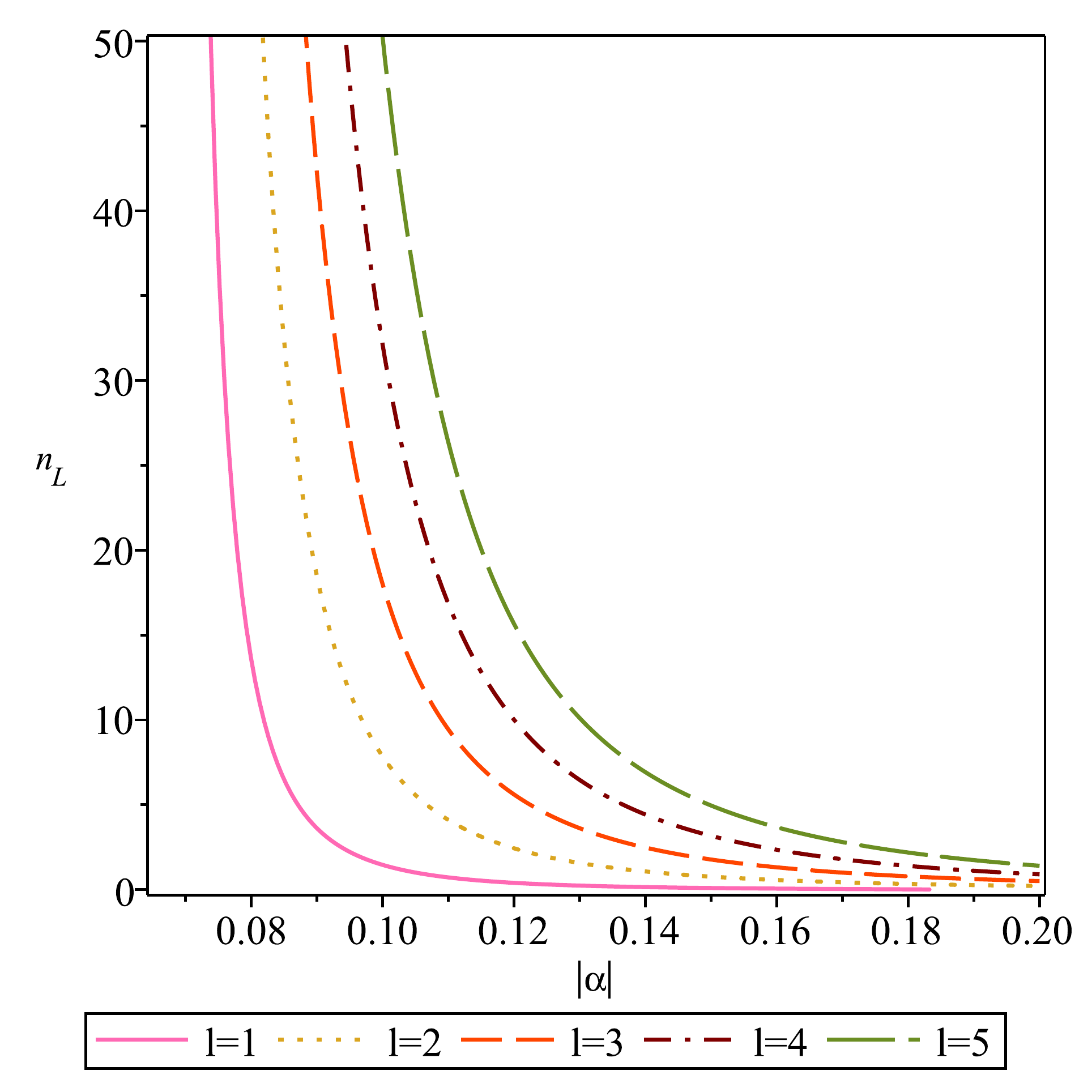}
		\caption{}
		\label{fig:3g}
	\end{subfigure}
	\caption{(a) and (b) $n_L$ as a function of $M$, with $\Omega=1,~l=1,~|\alpha|=0.05,~r_+=1$, and several numerical values of $Q$. (c) $n_L$ as a function of $\Omega$, with $Q=0.1,~l=1,~|\alpha|=0.05,~r_+=1$, and several numerical values of $M$. (d), (e), and (f) $n_L$ as a function of $Q$, with $M=3.5,~l=1,~|\alpha|=0.05,~r_+=1$, and several numerical values of $\Omega$. (g) $n_L$ as a function of $|\alpha|$, with $Q=0.1,~\Omega=1,~M=3.5,~r_+=1$, and several numerical values of $l$.}
	\label{fig:3}
\end{figure}
respectively, and
\be
{n_R} = 0,\label{nz}
\ee
\be {n_L} = \frac{{{r_ * }^2{r_ + }K\left( {{\Omega ^2} - {\Xi ^2}{l^2}} \right)\sqrt {\beta {r_ + }{r_ * }\delta } }}{{2\Omega {l^2}\Xi {{\left( {{r_ + } + {r_ * }} \right)}^3}}}.\label{ns}
\ee
The constant $\delta$ in Eqs. (\ref{tempR})--(\ref{ns}), is given by $\delta  =  {3{\Omega ^2}{r_ + }^2 - {r_ + }{r_ * }\left( {K{\Xi ^2}{l^4} - 2{\Omega ^2}} \right) + {\Omega ^2}{r_ * }^2}$. 
Moreover, we find two constraints for the parameters of the black hole solutions (\ref{metric}), such as
\begin{equation}
\frac{{{\Omega ^2}{\Xi ^2}{l^4}{{\left[ { - 3{r_ + }^2 + {r_ * }{r_ + }\left( {K{l^2} - 2} \right) - {r_ * }^2} \right]}^2}}}{{{K^2}\beta {r_ + }^3{r_ * }{{\left( {{r_ + } - {r_ * }} \right)}^2}{{\left( {{\Xi ^2}{l^2} - {\Omega ^2}} \right)}^2}}} = \frac{{{l^4}{\Omega ^2}}}{{{r_ + }^2K{{\left( {{\Xi ^2}{l^2} - {\Omega ^2}} \right)}^2}{{\left( {{r_ + } - {r_ * }} \right)}^2}\beta }},\label{c1}
\end{equation}
and
\be
\frac{{{\Omega ^2}{\Xi ^2}{l^4}{{\left( {K{l^2}{r_ * }^3 - {r_ + }^3 - 2{r_ * }{r_ + }^2 - 3{r_ * }^2{r_ + }} \right)}^2}}}{{{K^2}\beta {r_ + }{r_ * }^5{{\left( {{r_ + } - {r_ * }} \right)}^2}{{\left( {{\Omega ^2} - {\Xi ^2}{l^2}} \right)}^2}\left( {K{l^4}{\Xi ^2}{r_ * }{r_ + } - 3{\Omega ^2}{r_ + }^2 - 2{\Omega ^2}{r_ * }{r_ + } - {\Omega ^2}{r_ * }^2} \right)}} = \frac{{{l^4}{\Omega ^2}}}{{{r_ + }^2K{{\left( {{\Xi ^2}{l^2} - {\Omega ^2}} \right)}^2}{{\left( {{r_ + } - {r_ * }} \right)}^2}\beta }}.\label{c2}
\ee
\begin{figure}[H]
	\centering
	\begin{subfigure}{0.325\linewidth}
		\includegraphics[width=\linewidth]{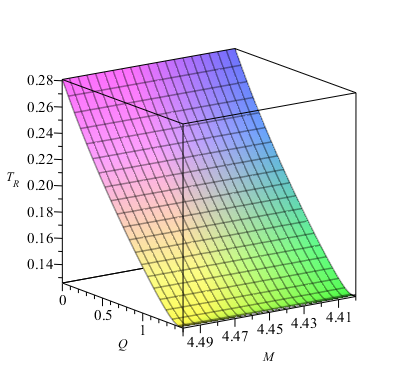}
		\caption{} 
		\label{fig:4a}
	\end{subfigure}
	\begin{subfigure}{0.325\linewidth}
		\includegraphics[width=\linewidth]{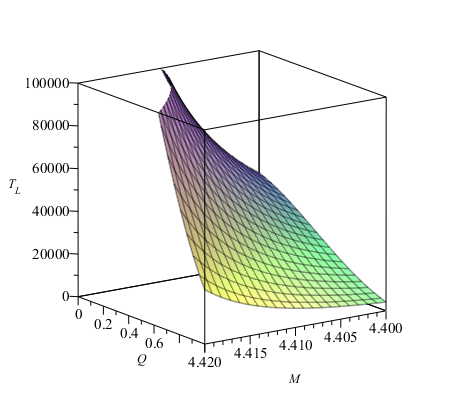}
		\caption{} 
		\label{fig:4b}
	\end{subfigure}
	\begin{subfigure}{0.325\linewidth}
		\includegraphics[width=\linewidth]{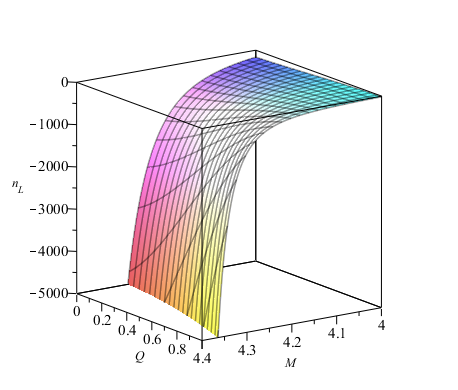}
		\caption{} 
		\label{fig:4c}
	\end{subfigure}
	\caption{(a) $T_R$ as a function of $M$ and $Q$. (b) $T_L$ as a function of $M$ and $Q$. (c) $n_L$ as a function of $M$ and $Q$. In these plots, we set in which, $\Omega=1,~l=1,~|\alpha|=0.05,~\text{and}~r_+=1$.}
	\label{fig:4}
\end{figure}
\end{widetext}
We note that Eqs. (\ref{c1}) and (\ref{c2}) restrict the black hole parameters, in accord with existence of the real positive values for the outer horizon (and any other horizons), as the roots of the triquadratic algebraic equation 
\be 
A(r)=0,
\ee 
where $A(r)$ is given in (\ref{Am}).\\
\indent Figure \ref{fig:1} shows the typical behavior of the right temperature (\ref{tempR}) versus different black hole parameters. Figures \ref{fig:1a}, \ref{fig:1b} and \ref{fig:1c} show the right temperature versus $M$, $\Omega$ and $|\alpha|$, respectively. Moreover, Figs. \ref{fig:1d}, \ref{fig:1e} and \ref{fig:1f} show the right temperature versus $Q$, for three different ranges of the electric charge, where the black hole (\ref{metric}) has real positive outer horizon.
\\
\indent Figure \ref{fig:2} shows the typical behavior of the left temperature (\ref{tempL}) versus different black hole parameters. Figures \ref{fig:2a}, \ref{fig:2b} and \ref{fig:2c} show the left temperature versus $M$, for three different ranges of the mass parameter, where the black hole (\ref{metric}) has real positive outer horizon. Figures \ref{fig:2d}, \ref{fig:2e} and \ref{fig:2f} show the left temperature versus $Q$, for three different ranges of the electric charge, where the black hole (\ref{metric}) has real positive outer horizon. Moreover, Fig. \ref{fig:2g} shows the left temperature versus $\Omega$. For several values of the mass parameter, the left temperature is positive definite. Finally, Fig. \ref{fig:2i} shows the monotonically increasing behavior for the left temperature, versus $\vert \alpha \vert$.
\\
\indent Moreover, Figs. \ref{fig:3a} to \ref{fig:3g} show the behavior of $n_L$ as in (\ref{ns}), versus the black hole parameters $M$, $\Omega$, $Q$ and $\vert \alpha \vert$. We note that $n_R=0$, according to Eq. (\ref{nz}).
\\
\indent We also plot the right temperature (\ref{tempR}), the left temperature (\ref{tempL}), and $n_L$ (\ref{ns}), versus the black hole mass parameter $M$ and the electric charge $Q$, in Figs. \ref{fig:4a}, \ref{fig:4b}, and \ref{fig:4c}, respectively.
\\
\indent We should emphasize that \slr~is a \textit{local} hidden symmetry, for the solution space of massless scalar field in the near region of the rotating charged AdS black holes (\ref{metric}), in quadratic $f(T)$ gravity. The \textit{local} hidden symmetry is generated by the vector fields (\ref{h1})--(\ref{hbm1}), which are not periodic under the angular identification
\be\label{periodic}
\phi\sim \phi +2\pi.
\ee
These symmetries cannot be used to generate new global solutions from the old ones. This can be interpreted as a statement that the \slr~symmetry is spontaneously broken down to $U\left(1\right)_L\times U\left(1\right)_R$ subgroup under the angular identification (\ref{periodic}). As a result, we can identify the left and right temperatures of the dual CFT.
\begin{widetext}

\begin{figure}[H]
	\centering
	\begin{subfigure}{0.3\linewidth}
		\includegraphics[width=\linewidth]{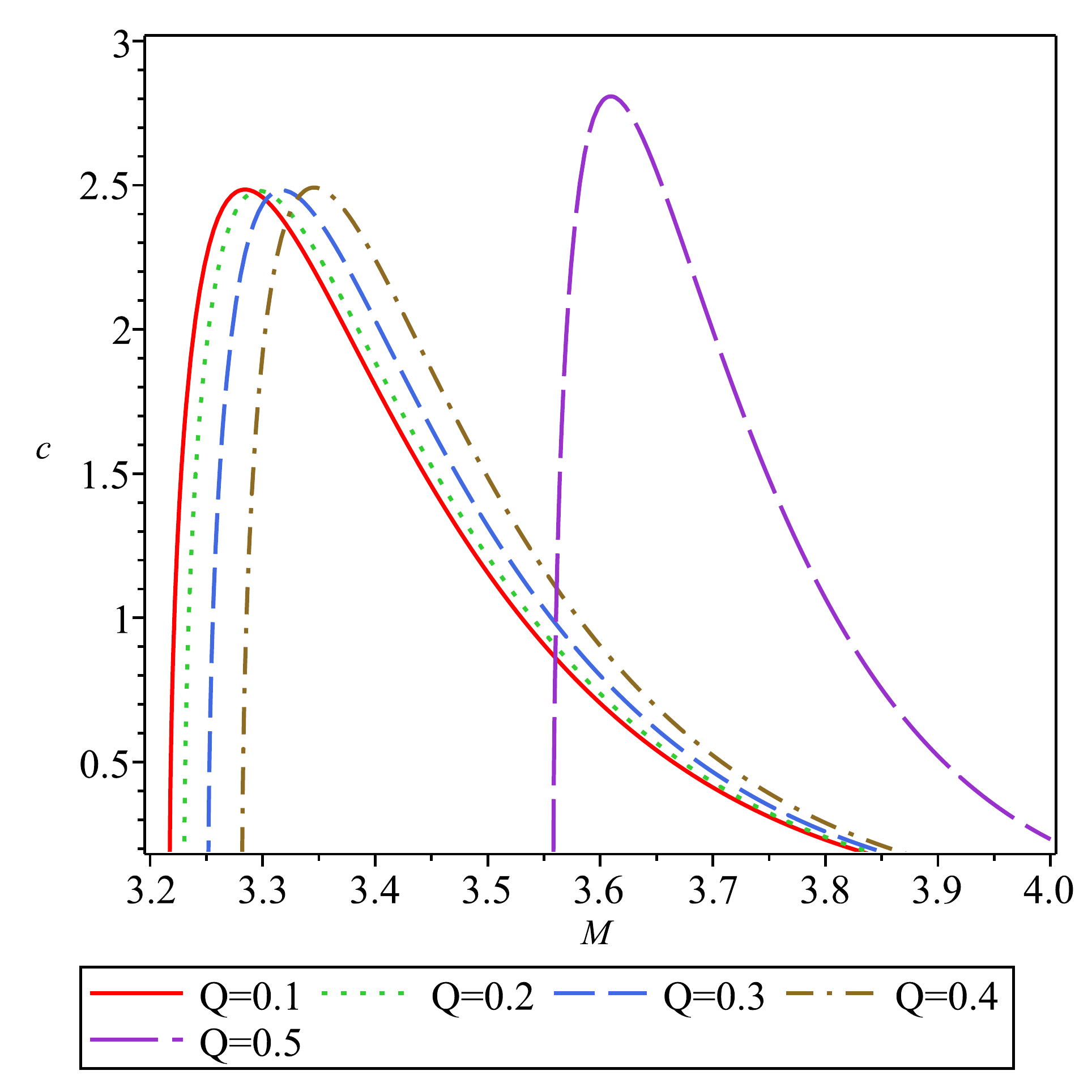}
		\caption{} 
		\label{fig:5a}
	\end{subfigure}
	\begin{subfigure}{0.3\linewidth}
		\includegraphics[width=\linewidth]{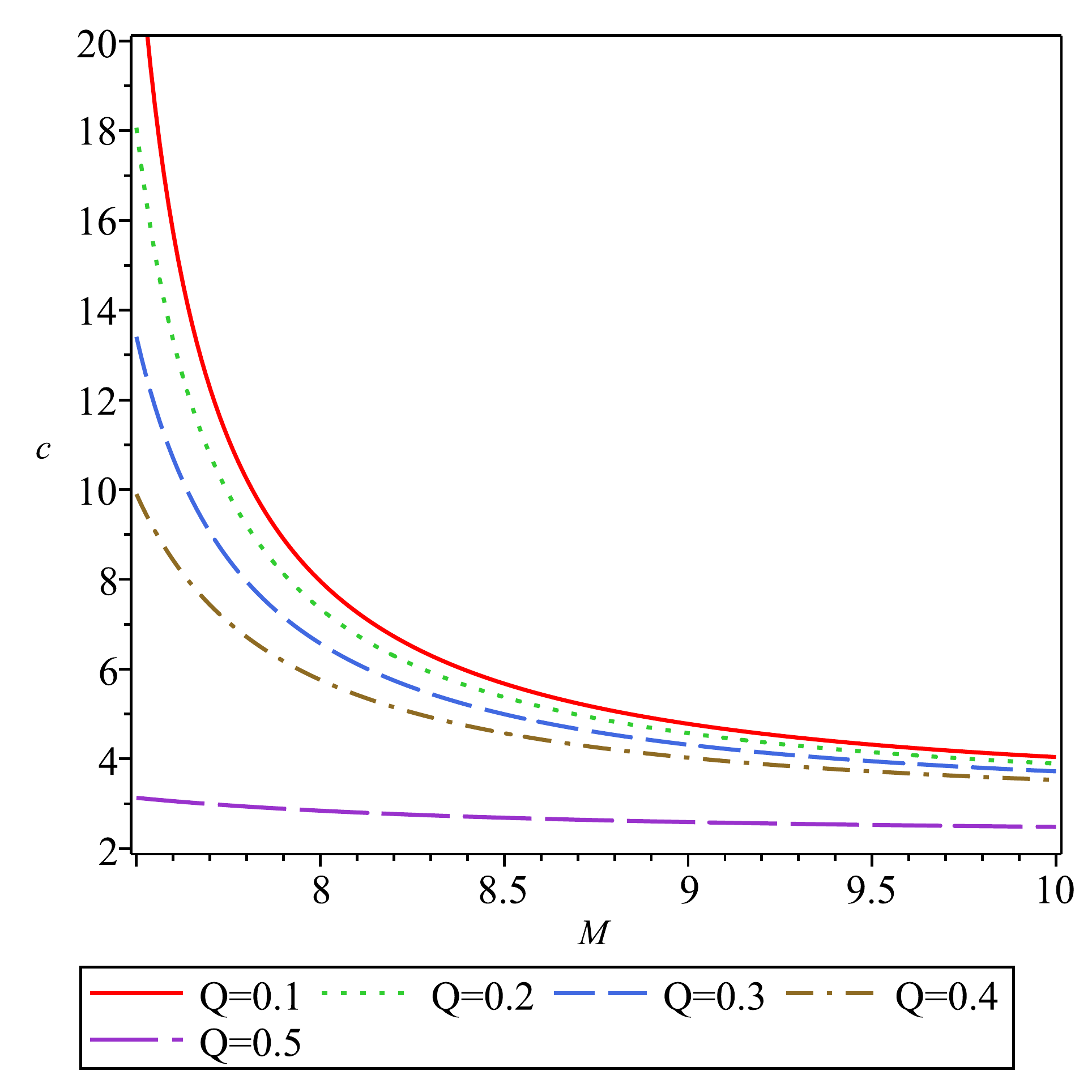}
		\caption{}
		\label{fig:5b}
	\end{subfigure}
	\begin{subfigure}{0.3\linewidth}
		\includegraphics[width=\linewidth]{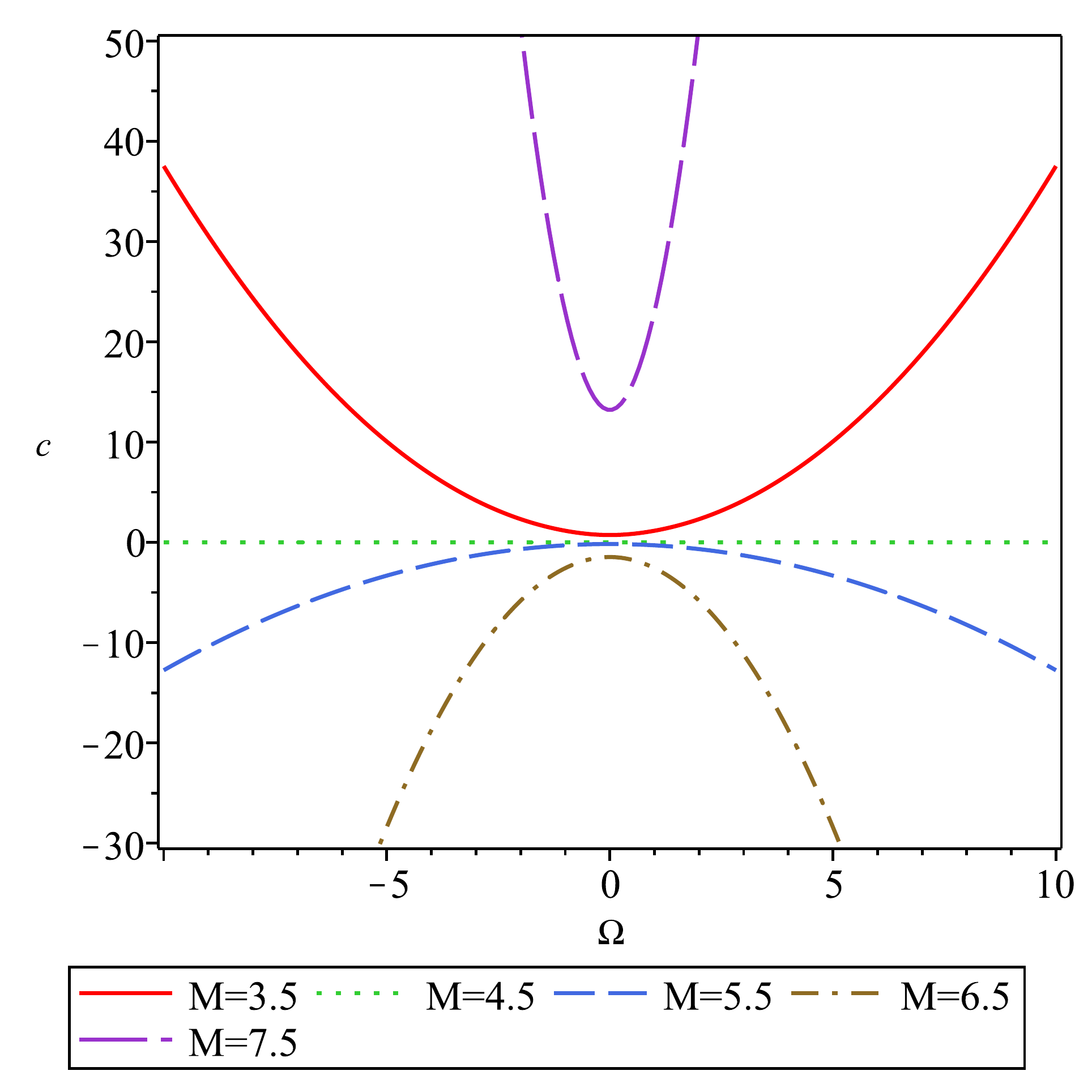}
		\caption{}
		\label{fig:5c}
	\end{subfigure}
	\begin{subfigure}{0.3\linewidth}
		\includegraphics[width=\linewidth]{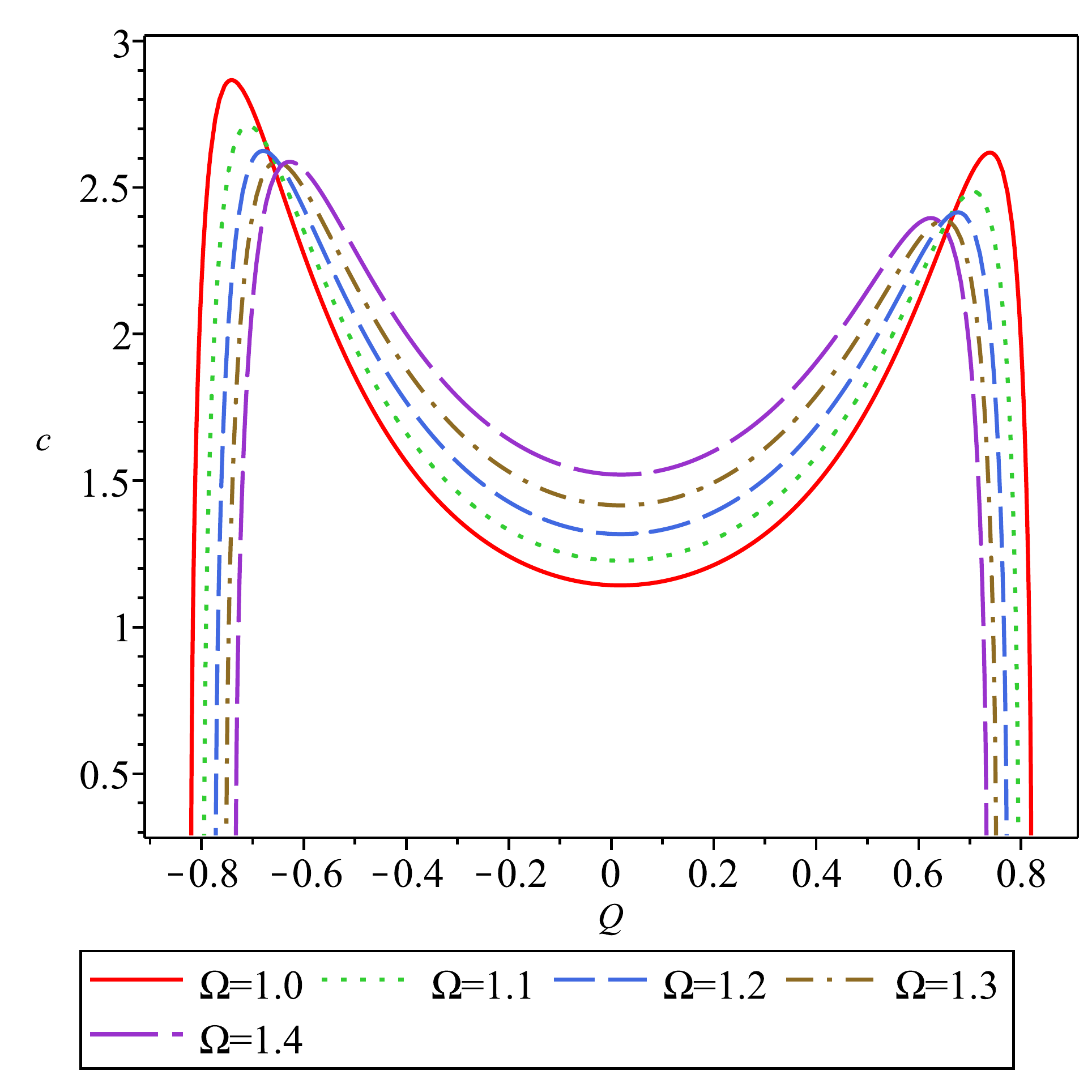}
		\caption{} 
		\label{fig:5d}
	\end{subfigure}
	\begin{subfigure}{0.3\linewidth}
		\includegraphics[width=\linewidth]{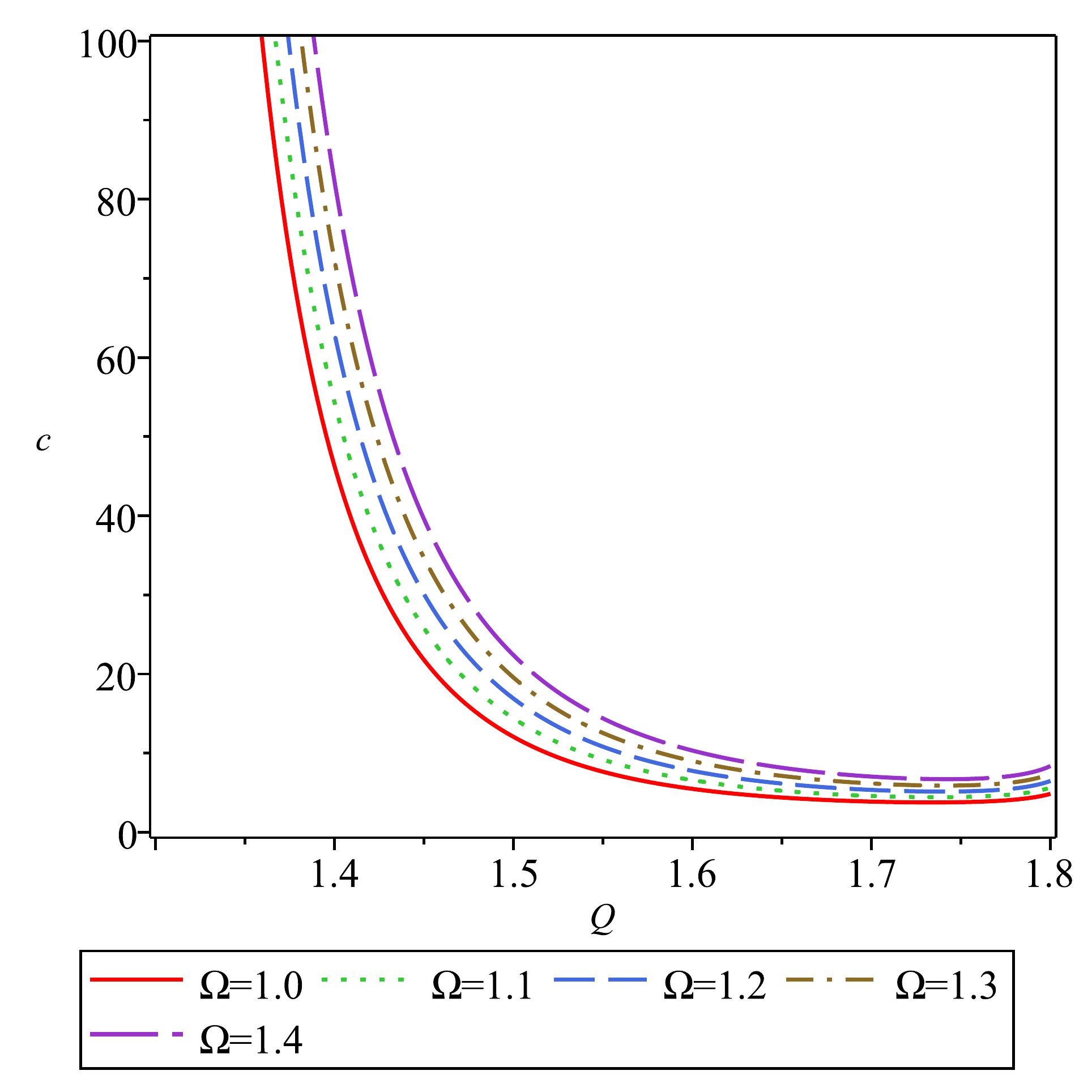}
		\caption{}
		\label{fig:5e}
	\end{subfigure}
	\begin{subfigure}{0.3\linewidth}
		\includegraphics[width=\linewidth]{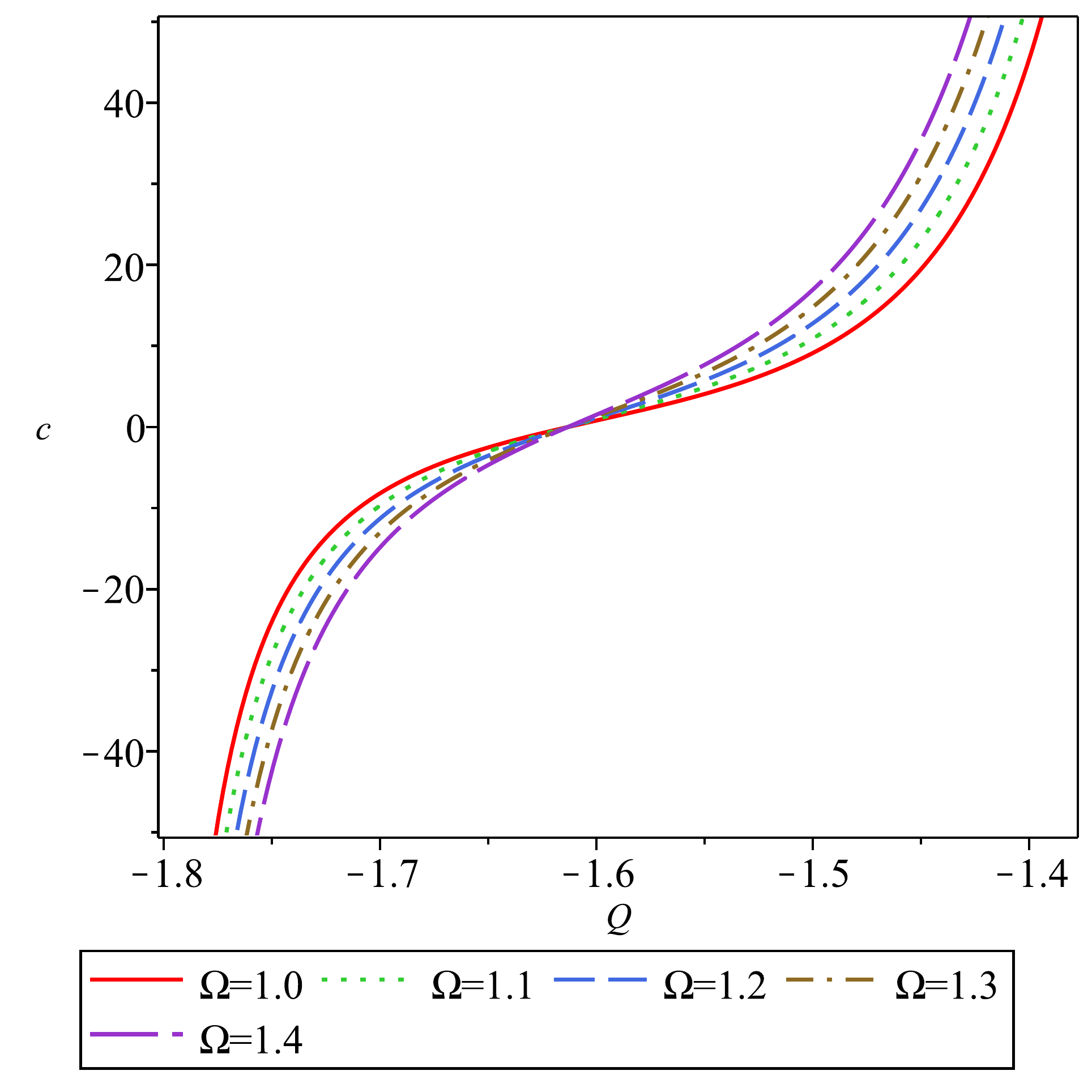}
		\caption{}
		\label{fig:5f}
	\end{subfigure}
	\begin{subfigure}{0.3\linewidth}
		\includegraphics[width=\linewidth]{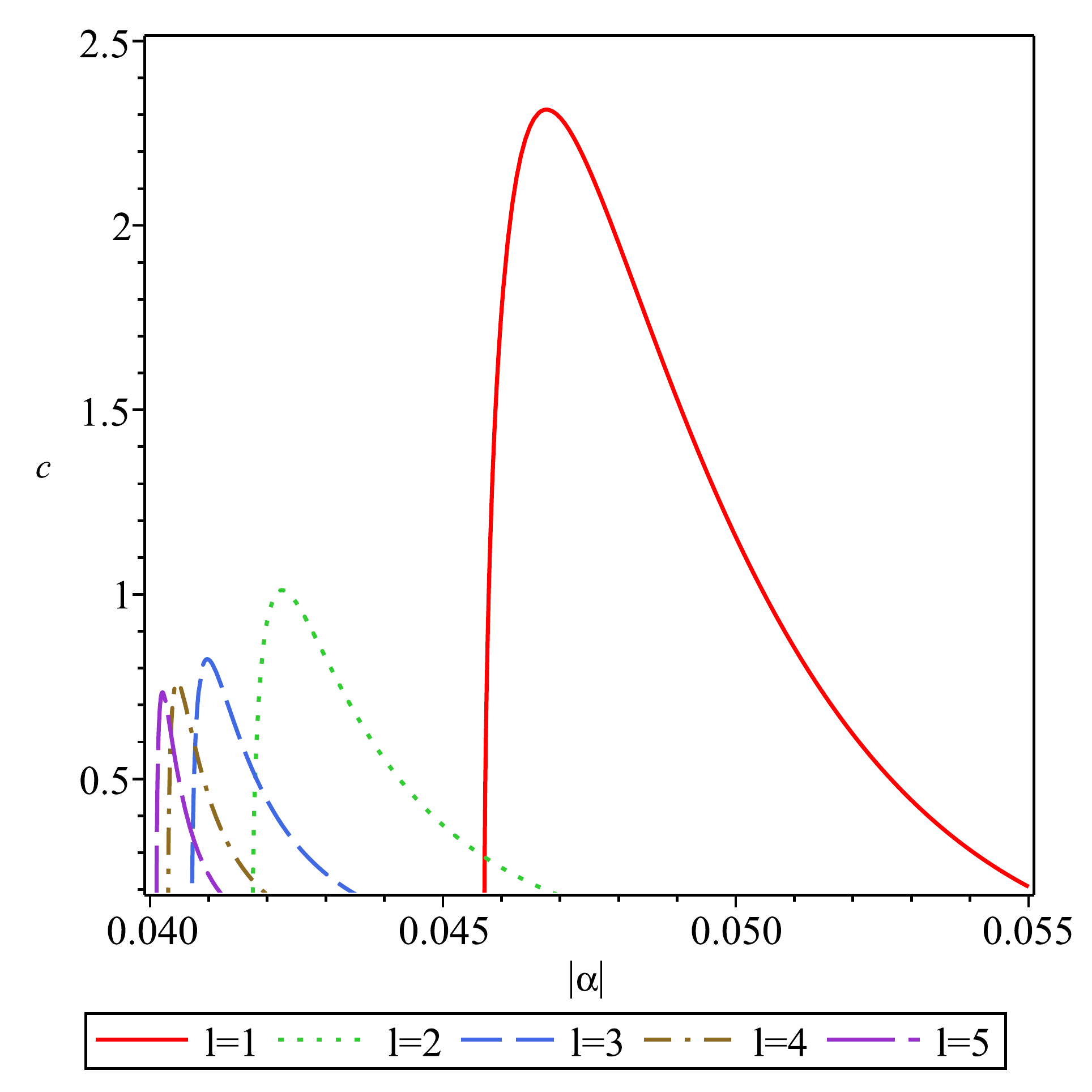}
		\caption{}
		\label{fig:5g}
	\end{subfigure}
	\caption{(a) and (b) $c$ as a function of $M$, with $\Omega=1,~l=1,~|\alpha|=0.05,~L=1,~r_+=1$, and several numerical values of $Q$. (c) $c$ as a function of $\Omega$, with $Q=0.1,~l=1,~|\alpha|=0.05,~L=1,~r_+=1$, and several numerical values of $M$. (d), (e), and (f) $c$ as a function of $Q$, with $M=3.5,~l=1,~|\alpha|=0.05,~L=1,~r_+=1$, and several numerical values of $\Omega$. (g) $c$ as a function of $|\alpha|$, with $\Omega=1,~Q=0.1,M~=3.5,~L=1,~r_+=1$, and several numerical values of $l$ }
	\label{fig:5}
\end{figure}

\end{widetext}
\section{CFT Entropy}
\begin{figure}[H]
	\centering
	\includegraphics[width=0.8\linewidth]{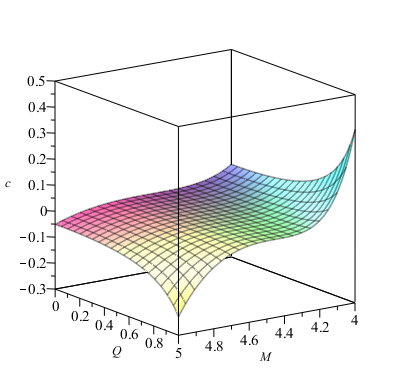}
	\caption{$c$ as a function of $M$ and $Q$. In these plots, we set in which, $\Omega=1,~l=1,~|\alpha|=0.05,~L=1,~\text{and}~r_+=1$.}
	\label{fig:6}
\end{figure}

We recall the Cardy entropy formula for the dual 2D CFT with temperatures $T_{L}$ and $T_{R}$
\be\label{cardy}
S_{CFT}=\frac{\pi^2}{3}\left(c_L T_L + c_R T_R\right),
\ee
where $c_L$ and $c_R$ are the corresponding central charges for the left and right sectors. The central charges can be derived from the \textit{asymptotic symmetry group} of the near-horizon (near-)extremal black hole geometry. There is no derivation for the central charges of the CFT dual to the nonextremal black holes, that we consider in this article. Of course, we expect that the conformal symmetry of the extremal black holes connects smoothly to those of the nonextremal black holes, for which the central charges are the same. The near-horizon extremal geometry for spacetime (\ref{metric}) is still unknown and it is not a straightforward task to find that due to the triquadratic behavior of the metric function $A(r)$. As a result, we turn the logic around and consider the favorite holographic situation, in which, the Cardy entropy (\ref{cardy}) produces exactly the macroscopic entropy (\ref{entropy1}). Substituting Eqs. (\ref{entropy2}), (\ref{tempR}), and (\ref{tempL}) to Eq. (\ref{cardy}), we find the central charges
\begin{widetext}
\be\label{cc}\nonumber
c \equiv c_L = c_R = \frac{{12\Xi \delta L\varpi {{\left( {{r_ + } + {r_ * }} \right)}^3}}}{{lK{r_ + }^2\left( {{\Xi ^2}{l^2} - {\Omega ^2}} \right)\left( {{r_ + }^3 + 2{r_ + }^2{r_ * } + 3{r_ + }{r_ * }^2 - {l^2}K{r_ * }^3} \right){{\left( {Q\sqrt {6\left| \alpha  \right|}  + {r_ + }^2} \right)}^2}\sqrt {\beta {r_ + }{r_ * }\delta } }},
\ee
where $\varpi  = {r_ + }^2\left( {{r_ + }^2Q\sqrt {6\left| \alpha  \right|} /2 + 7{r_ + }^4/18 + M\left| \alpha  \right|{r_ + } - 3{Q^2}\left| \alpha  \right|} \right) - 7\sqrt {6{{\left| \alpha  \right|}^3}} {Q^3}/3$. 
\end{widetext}
We note that we only consider CFTs, in which the left and right central charges are equal, $c\equiv c_{L}=c_{R}$ \cite{ElShowk:2011ag,Compere:2012jk}. In Figs. \ref{fig:5}--\ref{fig:6} we plot the behavior of the central charges (\ref{cc}) of the dual CFT, as a function of $M$, $\Omega$, $Q$, and $|\alpha|$.
\section{Conclusions}
\indent In this paper, we extend the concept of black hole holography to the nonextremal 4D rotating charged AdS black holes in $f$($T$)-Maxwell theory with a negative cosmological constant. We explicitly construct the hidden conformal symmetry for the rotating black holes in $f$($T$)-Maxwell theory with a negative cosmological constant. We mainly consider the near-horizon region, as the metric function which determines the event horizon, is a triple-quadratic equation. In this region, we show that the radial equation of the scalar wave function could be written as the \slr~ squared Casimir equation, indicating a \textit{local} hidden conformal symmetry acting on the solution space. The conformal symmetry is spontaneously broken under the angular identification $\phi\sim\phi+2\pi$, which suggests the rotating charged AdS black holes in quadratic $f$($T$) gravity, should be dual to the  finite temperatures $T_L$ and $T_R$ mixed state, in the 2D CFT. Instead of calculating the central charges using the \textit{asymptotic symmetry group}, we calculated the central charges by assuming the Cardy entropy for the dual CFT, matches the macroscopic Bekenstein-Hawking entropy. These results suggest that rotating charged AdS black holes in quadratic $f$($T$) gravity with particular values of $M$, $\Omega$, $Q$, and $|\alpha|$, are dual to a 2D CFT.\\
\indent It is an open question to find the near-horizon (near-)extremal geometry of the rotating charged AdS spacetime in quadratic $f$($T$) gravity. We may calculate the central charges using the \textit{asymptotic symmetry group} to confirm our results in this article. We can also study on various kinds of superradiant scattering off the near-extremal black hole as a further evidence to support the holographic picture for the nonextremal 4D rotating charged AdS black holes in $f$($T$)-Maxwell theory with a negative cosmological constant. We leave addressing these open questions for future articles.
\begin{acknowledgments}
This work is supported by the Natural Sciences and Engineering Research Council of Canada. C. B. is also supported by the William Rowles Fellowship in Physics and Engineering Physics, University of Saskatchewan.
\end{acknowledgments}   
\bibliography{apssamp}
\end{document}